\documentstyle[eqsecnum,preprint,12pt,aps,epsf]{revtex}

\begin{document}

\preprint{\begin{minipage}{3cm}\begin{flushright}
UTHEP-383\\\vspace*{-0.4cm}UTCCP-P-39
\end{flushright}\end{minipage}}
\title{Two-dimensional Lattice Gross-Neveu Model \\
with Wilson Fermion Action \\
at Finite Temperature and Chemical Potential}
\author{Taku Izubuchi$^{a)}$, Junichi Noaki$^{a)}$ and 
Akira Ukawa$^{a,b)}$}
\address{Institute of Physics$^{a)}$,  University of Tsukuba\\
        Tsukuba, Ibaraki 305-8571, Japan\\\ \\
        Center for Computational Physics$^{b)}$\\
        University of Tsukuba, Tsukuba, Ibaraki 305-8571, Japan}
\date{May 21, 1998}
\maketitle
\begin{abstract}
We investigate the phase structure of the two-dimensional lattice
Gross-Neveu model formulated with the Wilson fermion action to leading 
order of $1/N$ expansion.  Structural change of the parity-broken phase under 
the influence of finite temperature and chemical  potential is studied. 
The connection between the lattice phase structure and the chiral phase 
transition of the continuum theory is clarified. 
\end{abstract}
\newpage

\section{Introduction}
In spite of significant effort expended over the years\cite{wilsonreview}, 
thermodynamic studies of lattice QCD with the Wilson fermion action 
have been lagging behind those with the Kogut-Susskind action.  
The origin of difficulty is the explicit breaking of chiral 
symmetry due to the Wilson term, with ensuing complications in analysis 
of chiral properties.  It has become clear only recently that 
the finite-temperature phase diagram has an unconventional 
structure\cite{AUU}, 
having a region of spontaneously broken parity and flavor 
symmetry\cite{Aoki1,Aoki2,Sharpe,Bitar} 
in addition to the usual parity-flavor symmetric phase.  
While this development has considerably clarified a number of puzzling 
features observed in numerical simulations 
in the past\cite{wilsonreview}, 
there are still questions needing further elucidation. 
One of the questions is how the continuum limit is to be taken with the new 
phase diagram.  The parity-flavor broken phase has an extension which depends 
on the temporal lattice size, and the tuning of parameters necessary to 
achieve a continuum limit having chiral symmetry has not been explored 
in detail.
Another interesting question is how the phase diagram generalizes for finite 
quark chemical potential corresponding to finite baryon density.  
As is well known, finite-density studies of lattice QCD have been 
plagued with serious problems\cite{Barbour}. 
A conceptual understanding of the phase diagram is a prerequisite in 
numerical studies of this difficult problem. 

In this article we carry out an analytical exploration of these problems 
in the context of the two-dimensional Gross-Neveu model\cite{Gross-Neveu} 
on the lattice formulated with the 
Wilson fermion action\cite{Eguchi-Nakayama,Aoki-Higashijima}.  
The model shares with QCD the feature of asymptotic freedom and 
spontaneously broken chiral symmetry. 
Furthermore it is analytically solvable in $1/N$ expansion.  
These points make the model an useful arena for exploration of theoretical 
issues with lattice QCD thermodynamics with the Wilson quark action. 
Indeed the model has provided significant hints for understanding 
the structure of the finite-temperature phase diagram\cite{AUU}.

In the continuum the phase diagram of the Gross-Neveu model 
on the plane of temperature 
$T$ and fermion chemical potential $\mu$ was determined 
some time ago\cite{Wolff}.  In the leading order of $1/N$ expansion, 
the $(T,\mu)$ plane is divided into two phases, a chirally broken
phase at low temperatures and small chemical potential and a symmetric phase 
at high temperatures and large chemical potential, 
separated by a phase boundary. 
Along the phase boundary, the transition is of second order for 
small chemical potential, which 
however, changes into a first-order transition for large chemical potential. 
Our aim in this article will be, first, to determine the phase diagram 
of the lattice model for finite temporal lattice sizes corresponding to
finite temperature and finite chemical potential, and, second, to study 
how the continuum phase diagram is recovered as one takes 
the limit of continuum space-time. 

This paper is organized as follows. 
In Sec.~\ref{sec:two}, 
after a brief review of the Gross-Neveu model in the continuum,   
we formulate the lattice model with the Wilson fermion action at 
finite temperature and chemical potential, 
and analyze the behavior of the effective
potential toward the continuum limit.
In Sec.~\ref{sec:three} 
the phase structure of the model at zero temperature and zero chemical 
potential is studied.
In Sec.~\ref{sec:four} 
effects of finite temporal lattice size ({\it i.e.,} of 
finite temperature) on the phase diagram is examined, 
and the continuum extrapolation is studied. 
The case of finite chemical potential is treated in Sec.~\ref{sec:five}
 where we 
consider in detail how the difference in the order of phase transition 
observed in the continuum theory arises in the context of the lattice 
model. We conclude with a summary in Sec.~\ref{sec:six}. 
 
\section{Analytical Examinations}
\label{sec:two}

\subsection{Continuum Theory}
The Gross-Neveu model in two-dimensional Euclidean continuum space-time
is defined by the Lagrangian density,
\begin{eqnarray}
{\mathcal L}=\bar{\psi}(\gamma_\mu\partial_\mu+m)\psi-\frac{g^2}{2N}
[(\bar{\psi}\psi)^2+(\bar{\psi}i\gamma_5\psi)^2], \label{2}
\end{eqnarray}
where $\psi$ is an $N$-component spinor field and $g^2$ denotes the coupling 
constant.
Our convention for the two-dimensional $\gamma$ matrices is
\begin{eqnarray}
 \gamma_1=\sigma_2,\ 
\gamma_2=\sigma_1,\ \gamma_5=\sigma_3=i\gamma_1\gamma_2.
\end{eqnarray}
For massless fermion $m=0$, the model possesses $U(1)$ chiral symmetry 
defined by 
\begin{eqnarray}
\psi\to e^{i\theta\gamma_5}\psi\ ,\ 
\bar{\psi}\to e^{-i\theta\gamma_5}\bar{\psi}. \label{20}
\end{eqnarray}
In terms of the bosonic fields introduced by 
\begin{eqnarray}
\sigma\equiv -\frac{g^2}{N}\bar{\psi}\psi\ ,\ 
\pi\equiv -\frac{g^2}{N}\bar{\psi}i\gamma_5\psi\ , \label{3} 
\end{eqnarray}
chiral transformation (\ref{20}) represents a rotation on the
$\sigma$-$\pi$ plane by an angle $2\theta$. 

Statistical properties of the system at finite temperature $T$ can be 
examined by restricting the imaginary time extent of the space-time to 
$\beta=1/T$, and replacing energy integrals for fermions
by Matsubara mode sums over half-integer values according to
\begin{eqnarray}
k_2\to\omega_n=\frac{2\pi}{\beta}(n+1/2)\ \ ,\ n\in \bf{Z}.
\end{eqnarray}
In order to describe the system at finite chemical potential $\mu$,
we add $\mu\psi^\dagger\psi$ to ${\mathcal L}$ where $\psi^\dagger\psi$ 
is the fermion number operator.  These replacements, together with 
the introduction of the effective fields (\ref{3}), leads to the 
action given by
\begin{eqnarray}
S=\int_{-\infty}^{+\infty}\frac{dx_1}{2\pi}\int_{0}^{\beta}
\frac{dx_2}{2\pi}\Big[\bar{\psi}(\gamma_\mu\partial_\mu-m-\sigma
-i\gamma_5\pi+\mu\gamma_2)\psi-\frac{N}{2g^2}(\sigma^2+\pi^2)\Big].\label{4}
\end{eqnarray}

To leading order of $1/N$ expansion, the ground state of the model is 
determined by the minimum of the effective potential $V(\sigma, \pi)$ for 
a constant $\sigma$ and $\pi$ given by 
\begin{eqnarray}
V(\sigma,\,\pi)=\frac{1}{2g^2}(\sigma^2+\pi^2)-\frac{1}{\beta}
\sum_{n=-n_{\rm max}}^{n_{\rm max}-1}\int_{-M}^{+M}\frac{dk}{2\pi}\ln
[(\sigma+m)^2+\pi^2+k^2+(\omega_n+i\mu)^2], \label{8}
\end{eqnarray}
where $M$ represents an ultraviolet cutoff 
and $n_{\rm max}\equiv\beta M/(2\pi)$.
In the chiral limit $m\to 0$, $V$ is a function of $s^2\equiv\sigma^2+\pi^2$.
At zero temperature and chemical potential, the minimum of the 
effective potential is located at $s=\Lambda\equiv 2Me^{-\pi/g^2}\neq 0$, 
which signals spontaneous breakdown of chiral 
symmetry.  As the cutoff $M$ is removed $M\to\infty$, $g^2$ must converge 
to zero to keep $\Lambda$ finite, that is, the model is asymptotically free. 

\begin{figure}
\epsfxsize=0.9\textwidth
{\centering\epsfbox{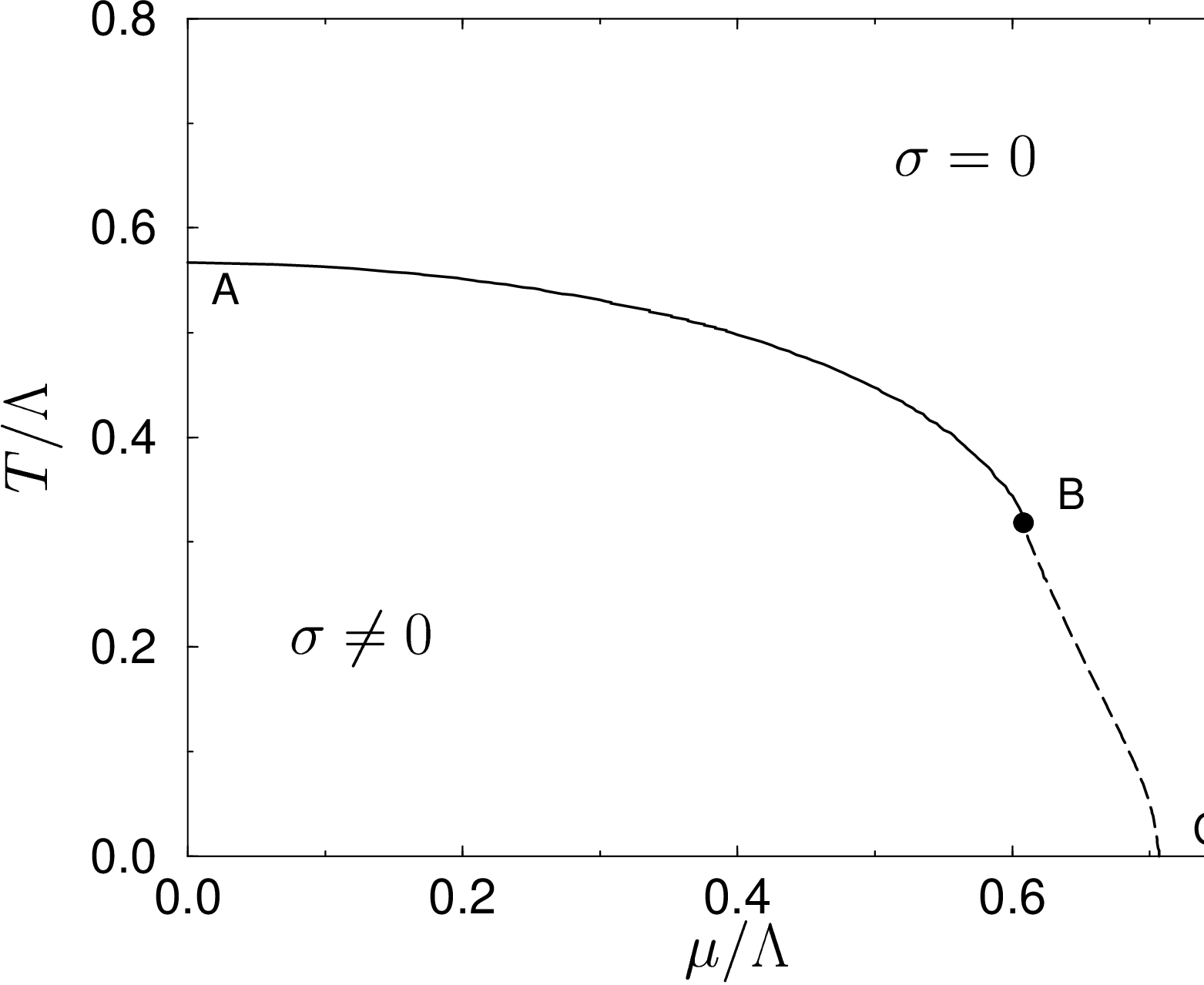} \par}

\vspace*{2cm}
\caption{
Phase diagram of the continuum Gross-Neveu model in the 
\protect\( (\mu/\Lambda ,T/\Lambda)\protect \)
parameter space. Solid and dashed curves represent a second- and a first-order 
segment of phase transition restoring chiral symmetry. 
\label{fig:CLPhase}}
\end{figure}

At finite temperature and chemical potential, Wolff \cite{Wolff} analyzed 
the effective potential (\ref{8}) and determined the phase diagram  
in the $T$-$\mu$ plane, which is reproduced in Fig.~\ref{fig:CLPhase}. 
Inside the boundary ABC chiral symmetry is spontaneously broken, while 
the region outside of this line is chirally symmetric.  The point $B$ is 
a tricritical point separating a second-order phase
boundary AB from a first-order one BC.

\subsection{Lattice Theory with Wilson Fermion Action}
Consider a two-dimensional lattice of a lattice spacing $a$.  
The lattice Gross-Neveu model with the Wilson fermion action is defined
by\cite{Eguchi-Nakayama}
\begin{eqnarray}
        \mathcal{L}_{\rm lat}&=&-\frac{1}{2a}\sum_{\mu}
         [\, \bar{\psi}(x)\,(r-\gamma_\nu)\,\psi(x+\hat{\mu})+\bar{\psi}(x)\,
        (r+\gamma_\mu)\,\psi(x-\hat{\mu})\,] \nonumber \\
              & &
         +\frac{1}{a}(2r+\delta ma)\bar{\psi}(x)\psi(x)-\frac{1}{2N}[\,
         g_\sigma^2(\bar{\psi}(x)\psi(x))^2+{g_\pi^2}
        (\bar{\psi}(x)i\gamma_5\psi(x))^2\,]. \label{10}
\end{eqnarray}
where $\delta m$ is a mass counterterm.  
We take the coupling constant $g_\sigma^2$ and $ g_\pi^2$ 
for the two four-fermion interaction terms 
to be different\cite{Aoki-Higashijima}, whose reason will become clear below. 
Hereafter we let Wilson parameter $r$ be unity.

Similar to the continuum theory, we define a pair of bosonic fields 
according to 
\begin{eqnarray}
\sigma\equiv -\frac{g_\sigma^2}{N}\bar{\psi}\psi+\delta m\ ,\ 
\pi\equiv -\frac{g_\pi{}^2}{N}\bar{\psi}i\gamma_5\psi\ , \label{30} 
\end{eqnarray}
In order to consider the system at finite temperature $T$, 
we take a lattice with $N_T=1/(Ta)$ sites in the temporal direction.  
A finite chemical potential $\mu$ is introduced through a modification 
of the hopping factor in the temporal direction\cite{chemical}. 
The Lagrangian of the model then takes the form, 
\begin{eqnarray}
        \mathcal{L_{\rm lat}}&\to&-\frac{1}{2a}[\,\bar{\psi}(x)\,
         (1-\gamma_1)\,
        \psi(x+\hat{1})+\bar{\psi}(x)\,(1+\gamma_1)\,\psi(x-\hat{1})\,]
        \nonumber \\
        & &-\frac{1}{2a}[\,\bar{\psi}(x)\,(1-\gamma_2)\,e^{-\mu a}\,\psi(x+
        \hat{2})+\bar{\psi}(x)\,(1+\gamma_2)\,e\,^{\mu a} \,
        \psi(x-\hat{2})\,] 
        \nonumber \\    
        & &+\frac{1}{a}(2+\delta ma)\bar{\psi}(x)\psi(x)
         -\frac{N}{2g_\sigma^2}(\sigma-\delta m)^2-\frac{N}{2g_\pi^2}
         \pi^2.  \label{160}
\end{eqnarray}

With these modifications, the effective potential $V_L = Va^2$ in lattice units
is given by 
\begin{eqnarray}
V_{L}(\sigma_L, \pi_L)&=&\frac{1}{2g_\sigma^2}(\sigma_L-\delta m_L)^2
         +\frac{1}{2g_\pi^2}\pi_L^2-\int_{-\pi}^{\pi}
        \frac{d\xi}{2\pi}\mathcal{J}(\xi)\ ,\label{170}
\end{eqnarray}
with 
\begin{eqnarray}
 {\mathcal J}(\xi)&\equiv&\frac{1}{N_T}\sum_{n=0}^{N_T-1}
         \ln[A-B\cos(\omega_na+i\mu_L)], \quad 
        \omega_na\equiv\frac{2\pi}{N_T}(n+\frac{1}{2}) ,\label{180}\\
 A&=&2+(\sigma_L+2)^2-2(\sigma_L+2)\cos\xi+\pi_L^2 , \label{190} \\
 B&=&2(\sigma_L+2)-2\cos\xi , \label{200}
\end{eqnarray}
where $\sigma_L\equiv \sigma a,\,\pi_L\equiv \pi a,\,
\delta m_L\equiv\delta ma$ and $\xi=k_1a$ are quantities in lattice units. 

The Matsubara mode sum in $\mathcal{J}(\xi)$ can be carried out as follows:
\begin{eqnarray}
\mathcal{J} (\xi)&=&\frac{1}{N_T}\ln\prod_{n=1}^{N_T}\Bigl[
A-\frac{B}{2}e^{-\mu_L-i\pi/N_T}e^{2\pi ni/N_T}-\frac{B}{2}e^{\mu_L+i\pi/N_T}
e^{-2\pi ni/N_T}\Bigl] \nonumber\\
&=&\frac{1}{N_T}\ln\Bigl[\Bigl(\frac{B}{2}\Bigl)^{N_T}e^{EN_T}
\prod_{n=1}^{N_T}(1-e^{-E-\mu_L-i\pi/N_T}e^{2\pi ni/N_T})\nonumber \\
& &\ \ \ \ \ \ \ \ \ \ \ \ \ \ \ \ \ \ \ \ \ 
\times\prod_{n=1}^{N_T}(1-e^{-E+\mu_L+i\pi/N_T}e^{-2\pi ni/N_T})\Bigl]
\nonumber\\
&=&\ln\frac{|B|}{2}+\mu_L+\frac{1}{N_T}\ln|1+e^{(E-\mu_L)N_T}|
+\frac{1}{N_T}\ln|1+e^{-(E+\mu_L)N_T}|\label{230},
\end{eqnarray}
where
\begin{eqnarray}
E\equiv\left\{
   \begin{array}{ll}
    \cosh^{-1}(A/B) & \mbox{if}\ A/B> 0\\
    \cosh^{-1}\bigl|A/B\bigl|+i\pi & \mbox{if}\ A/B<0\ ,
   \end{array}
\right.\label{240}
\end{eqnarray}
and we have used the formula
\begin{eqnarray}
\prod_{n=1}^N(1+xe^{2\pi ni/N})=1-(-x)^N.
\end{eqnarray}
The real part of $E$ has the meaning of the energy of the fermion one-particle
state for a given value of $\sigma_L\,,\,\pi_L$ and dimensionless space 
momentum $\xi$.
 
\subsection{Continuum Limit}
The lattice Gross-Neveu model as we defined above explicitly breaks chiral 
symmetry due to the Wilson term.  Toward the continuum limit $a\to 0$, 
an examination of effects of symmetry breaking is possible 
through an expansion of the effective potential in powers of 
$a$\cite{Aoki-Higashijima}.  We briefly recapitulate the analysis here 
as it raises an important point for the study of phase diagram 
in the following sections. 

We consider the effective potential (\ref{170}) at $T=0 (N_T=\infty)$ and 
$\mu=0$, and write it as 
\begin{eqnarray}
V_{L}(\sigma_L,\,\pi_L)&=&
\frac{1}{2g_\sigma^2}(\sigma_L-\delta m_L)^2+\frac{1}{2g_\pi^2}\pi_L^2 
-\int_{-\pi}^{+\pi}\frac{d^2 \xi}{(2\pi)^2}
\ln\left[D_0(\xi)+D_1(\xi)\right], \label{51}
\end{eqnarray}
where 
\begin{eqnarray}
D_0(\xi)&=& \sum_{\nu=1,2} \sin^2 \xi_\nu 
        +\left(\sum_{\nu=1,2}(1-\cos \xi_\nu ) \right)^2+\sigma_L^2+\pi_L^2,\\
D_1(\xi)&=&2\sigma_L\sum_{\nu=1,2}(1-\cos \xi_\nu ). 
\end{eqnarray}
The continuum limit of the effective potential is determined by terms 
of ${\cal O}(a^2)$ with $\sigma_L$ and $\pi_L$ regarded as 
${\cal O}(a)$. 
Since $D_1(\xi)={\cal O}(a)$ while $D_0(\xi)={\cal O}(1)$, we make an 
expansion in terms of $D_1(\xi)$ in (\ref{51}) obtaining 
\begin{eqnarray}
V_{L}(\sigma_L,\,\pi_L)&=&\frac{1}{2g_\sigma^2}(\sigma_L-
         \delta m_L)^2+\frac{1}{2g_\pi^2}\pi_L^2 
        -\int_{-\pi}^{+\pi}\frac{d^2 \xi}{(2\pi)^2}\ln D_0(\xi),\nonumber\\
&+&\sum_{n=1}^\infty \frac{(-1)^n}{n}
        \int_{-\pi}^{+\pi}\frac{d^2 \xi}{(2\pi)^2}
        \frac{D_1(\xi)^n}{D_0(\xi)^n}.\label{52}
\end{eqnarray}

The integral over $\ln D_0(\xi)$ in (\ref{52}) 
can be estimated by an expansion 
around $\sigma_L^2+\pi_L^2=0$, paying attention to logarithmic divergence of 
the first derivative of the integral with respect to $\sigma_L^2+\pi_L^2$. 
The result reads
\begin{eqnarray}
\int_{-\pi}^{\pi}\frac{d^2 \xi}{(2\pi)^2}\ln D_0(\xi)
&=&C_0(\sigma_L^2+\pi_L^2)
-\frac{1}{4\pi}(\sigma_L^2+\pi_L{}^2)\ln\frac{\sigma_L^2+\pi_L^2}{e}
+{\mathcal O}(a^4),\label{54}
\end{eqnarray} 
where
\begin{eqnarray}
C_0= 0.220634\cdots .
\end{eqnarray}

In the sum over $n$ in (\ref{52}) only the terms up to and including $n=2$
contribute to ${\cal O}(a^2)$, and $\sigma_L^2+\pi_L^2$
in $D_0(\xi)$ can be set to zero in this limit.  We then find that
\begin{eqnarray}
\int_{-\pi}^{+\pi}\frac{d^2 \xi}{(2\pi)^2}\frac{D_1(\xi)}{D_0(\xi)}
&=&2C_1\sigma_L+{\cal O}(a^3),\qquad 
C_1=\frac{2\sqrt{3}}{9},\label{53}\\
\int_{-\pi}^{+\pi}\frac{d^2 \xi}{(2\pi)^2}\frac{D_1(\xi)^2}{D_0(\xi)^2}
&=&4C_2\sigma_L^2+{\cal O}(a^4),\qquad 
C_2=\frac{2\sqrt{3}}{27}+\frac{1}{12\pi}. \label{533}
\end{eqnarray} 
The integrals (\ref{53}-\ref{533}) clearly show that the Wilson term 
distorts the effective potential in the $\sigma$ direction relative 
to that of $\pi$.  Collecting the results together we find
\begin{eqnarray}
V_{L}&=&-\left(\frac{\delta m_L}{g_\sigma^2}
+2C_1\right)\sigma_L+\left(\frac{1}{2g_\pi^2}-C_0\right)
 \pi_L^2+\left(\frac{1}{2g_\sigma^2}-C_0+2C_2\right)\sigma_L^2 
\nonumber \\
& &+\frac{1}{4\pi}(\sigma_L^2+\pi_L^2)\ln\frac{\sigma_L^2+\pi_L^2}{e}
+{\mathcal O}(a^3) \ ,\label{60}
\end{eqnarray}
The necessity for introduction of the two couplings $g_\sigma^2$ and $g_\pi^2$ 
should now be clear\cite{Aoki-Higashijima}: 
chiral symmetry would not be recovered 
toward the continuum limit unless one tunes the couplings.
The limit of massless fermion further requires a tuning of the mass 
parameter $\delta m_L$ to remove the linear term in $\sigma$.
A natural tuning will be
\begin{eqnarray}
\frac{1}{g_\pi^2}=\frac{1}{g_\sigma^2}+4C_2+{\cal O}(a), \label{110}
\end{eqnarray} 
and
\begin{eqnarray}
\frac{\delta m_L}{g_\sigma^2}&=&-2C_1+{\cal O}(a^2).\label{91}
\end{eqnarray}

Let us introduce the $\Lambda$ parameter through 
\begin{eqnarray}
 \Lambda \equiv \frac{c}{a}e^{-\pi /g_\sigma^2},
\qquad c\equiv e^{2\pi C_0-4\pi C_2}=0.57160\cdots . 
\label{65}
\end{eqnarray}
With the tuning (\ref{110}) this corresponds to a 
coupling renormalization given by 
\begin{eqnarray}
\frac{1}{2g_\sigma^2}&=&C_0-2C_2+\frac{1}{4\pi}
        \ln \frac{1}{\Lambda^2 a^2}, \label{70} \\
\frac{1}{2g_\pi^2}&=&C_0+\frac{1}{4\pi}\ln 
        \frac{1}{\Lambda^2 a^2}, \label{80} 
\end{eqnarray}
with which the effective potential (in physical units) 
takes the standard continuum form
\begin{eqnarray}
 V(\sigma,\,\pi)=\frac{1}{4\pi}(\sigma^2+\pi^2)
\ln(\frac{\sigma^2+\pi^2}
{e\Lambda^2})+{\mathcal O}(a). \label{100}
\end{eqnarray}

It is straightforward to extend the above analysis to the case of 
finite temperature and chemical potential.  No new divergences arise
in these cases, and the same tuning of parameters $g_\sigma^2, g_\pi^2$ and 
$\delta m_L$ as specified in (\ref{110}-\ref{91}) is necessary for the 
correct continuum limit. Hereafter, we often use this tuning ignoring the
${\cal O}(a)$ corrections. This is equivalent to taking 
\begin{eqnarray}
{\bf T}(g_\sigma^2) \equiv \left(-2C_1g_\sigma^2\,,
\,(1/g_\sigma^2+4C_2)^{-1} \right) \label{tunedT}
\end{eqnarray}
as a tuned point on the $(\delta m_L , g_\pi^2)$ plane.

We learn from the present analysis that 
understanding of the continuum limit requires an elucidation of the phase 
diagram of the lattice model in the three-dimensional parameter space 
of $(g_\sigma^2, g_\pi^2, \delta m_L)$.  This will be our basic viewpoint 
in the following sections. 

\section{Phase diagram at $T=0$ and $\mu=0$}
\label{sec:three}
We start our investigation of the phase structure of the lattice
model with the case of zero temperature and zero chemical potential. 
To leading order in $1/N$ expansion, the ground state of the model 
is determined by the pair of saddle point equations
\begin{eqnarray}
\frac{\partial V_{L}}{\partial \sigma_L}&=&
\frac{\sigma_L-\delta m_L}{g_\sigma^2}-
\left( \sigma_L F(\sigma_L, \pi_L)+G(\sigma_L, \pi_L)\right)=0,\label{124}\\
\frac{\partial V_{L}}{\partial \pi_L}&=&
\pi_L\left(\frac{1}{g_\pi^2}-F(\sigma_L, \pi_L)\right)=0, \label{125}
\end{eqnarray}
where 
\begin{eqnarray}
F(\sigma_L, \pi_L)&=&
\int_{-\pi}^{+\pi}\frac{d^2\xi}{(2\pi)^2}\frac{2}{\sum_\nu 
\sin^2\xi_\nu +(\sigma_L+\sum_\nu(1-\cos \xi_\nu ))^2+\pi_L^2}, \label{130}\\ 
G(\sigma_L, \pi_L)&=&
\int_{-\pi}^{+\pi}\frac{d^2\xi}{(2\pi)^2}\frac{2\sum_\nu
(1-\cos \xi_\nu )}{\sum_\nu\sin^2\xi_\nu +(\sigma_L+\sum_\nu
(1-\cos \xi_\nu ))^2+\pi_L^2} .\label{120}
\end{eqnarray}

The second equation allows a non-trivial solution with $\pi_L\ne 0$, 
which corresponds to spontaneous breakdown of parity symmetry, in 
addition to the parity-symmetric solution $\pi_L=0$. 
If the phase transition between the two phases is continuous, the phase 
boundary separating them can be determined by examining the limit of the 
parity-broken solution toward $\pi_L\to 0$.  This yields the pair of 
equations
\begin{eqnarray}
\frac{\sigma_L-\delta m_L}{g_\sigma^2}&=&
\sigma_L F(\sigma_L, 0)+G(\sigma_L, 0) , \label{121}\\
\frac{1}{g_\pi^2}&=&
F(\sigma_L, 0). \label{131} 
\end{eqnarray}
The second equation ensures that the pion mass $M_\pi$ 
vanishes along the phase boundary since, for $\pi_L=0$, $M_\pi$ is given by 
\begin{eqnarray}
\left(\frac{M_\pi}{\Lambda}\right)^2=\frac{\partial^2 V_L}{\partial \pi_L^2}
\Biggl|_{\pi_L=0}=\frac{1}{g_\pi^2}-F(\sigma_L, 0).
\label{150}
\end{eqnarray}
The structure of the phase boundary was previously studied for the 
case of an equal coupling $g_\sigma^2=g_\pi^2$ in Refs.~\cite{Aoki1,AUU}. 
We have to extend the analysis treating the couplings as independent.  

\begin{figure}
\epsfxsize=0.9\textwidth
{\centering \epsfbox{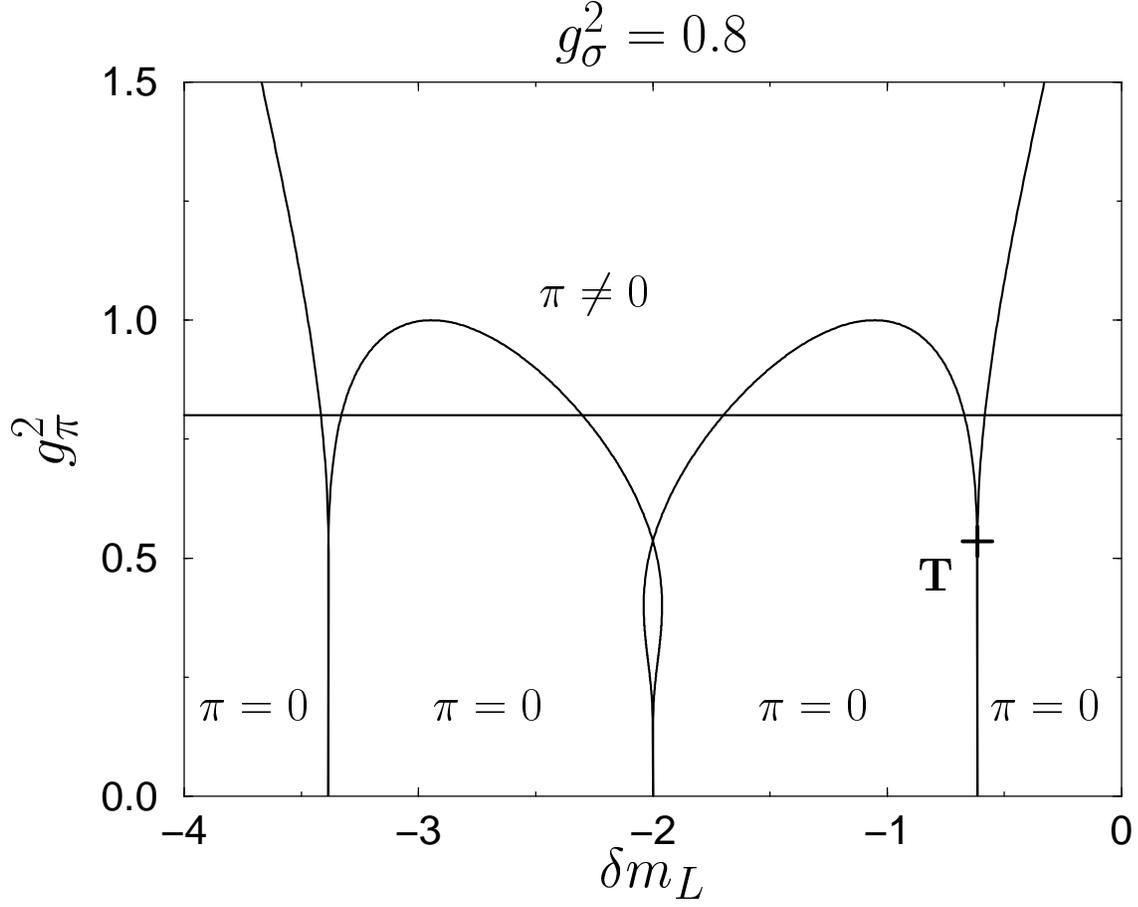} \par}

\vspace*{2cm}
\caption{Parity phase boundary for \protect\(g_\sigma^2=0.8\protect\)
on \protect\( (\delta m_L,g^{2}_{\pi })\protect \) plane. 
Detailed structure near the tuning point \protect\({\bf T}=
{\bf T}(g_\sigma^2)\protect\) for taking the continuum limit is shown 
in Fig.~\ref{fig:mu0PhaseDet}. 
\label{fig:mu0Phase}}
\end{figure}

The pair of equations (\ref{121}-\ref{131}) defines a surface in the 
three-dimensional parameter space $(g_\sigma^2, g_\pi^2, \delta m_L)$. 
We shall tentatively call this surface as {\it parity phase boundary}. 
In Fig.~\ref{fig:mu0Phase} we show the section of the surface on the 
$(\delta m_L, g_\pi^2)$ plane at $g_\sigma^2=0.8$.   
The region above the curve corresponds to the 
parity-breaking solution 
$\pi_L\ne 0$, while $\pi_L=0$ below it.  
As one moves along the curve from right to left, the value of $\sigma_L$ 
taken as 
the parameter of the curve decreases from positive to negative values.  
The bottom of the three valleys of the curve reaches $g_\pi^2=0$ at 
$\sigma_L=\sigma_{L0}=0, -2, $ and $-4$ from right to left, respectively, 
where $F(\sigma_L, 0)=+\infty$ and the mass parameter $\delta m_L$ equals
$\delta m_L=\delta m_{L0}(g_\sigma^2)\equiv 
\sigma_{L0}-g_\sigma^2 G(\sigma_{L0},0)$. 

The horizontal line marks the position of equal coupling $g_\sigma^2=g_\pi^2$.
We observe that there are three parity-broken intervals in between the 
parity-symmetric ones. Toward the weak-coupling limit $g_\sigma^2\to 0$, 
the horizontal line moves toward the bottom of the valley, which in turn 
converges toward $\delta m_{L0}(0)=\sigma_{L0}=0, -2, -4$. 
Hence each of the parity-broken intervals narrows to a point at
$g_\sigma^2=0$; the point at $\delta m_L=0$ is 
the conventional continuum limit,  while $\delta m_L=-2$ ($\delta m_L=-4$)
represents the point where the doublers with momentum $\xi=(0,\pi)$ and 
$(\pi, 0)$ ($\xi=(\pi, \pi)$) become massless physical modes. 
These are the results previously obtained in Refs.~\cite{Aoki1,AUU}.

As we have emphasized in the previous section, however, a proper continuum 
limit has to be taken along the line specified by (\ref{110}-\ref{91}). 
The point ${\bf T}(g_\sigma^2)$ corresponding to this line without 
${\cal O}(a)$ corrections, given in (\ref{tunedT}),  is marked by 
a cross in Fig.~\ref{fig:mu0Phase}.  
The parity-broken interval is significantly narrower near the point, and 
the location of the point itself is indistinguishable from the 
curve in the scale of the figure.  A more detailed study of the 
region near the point is clearly needed.  As we shall see, the phase 
structure near the point is far more complex than would seem from 
Fig.~\ref{fig:mu0Phase}. An indication is a self-crossing behavior of the 
curve already visible for the central valley in Fig.~\ref{fig:mu0Phase},  
which also occurs for the right and left valleys.

\begin{figure}
\epsfxsize=0.9\textwidth
{\centering \epsfbox{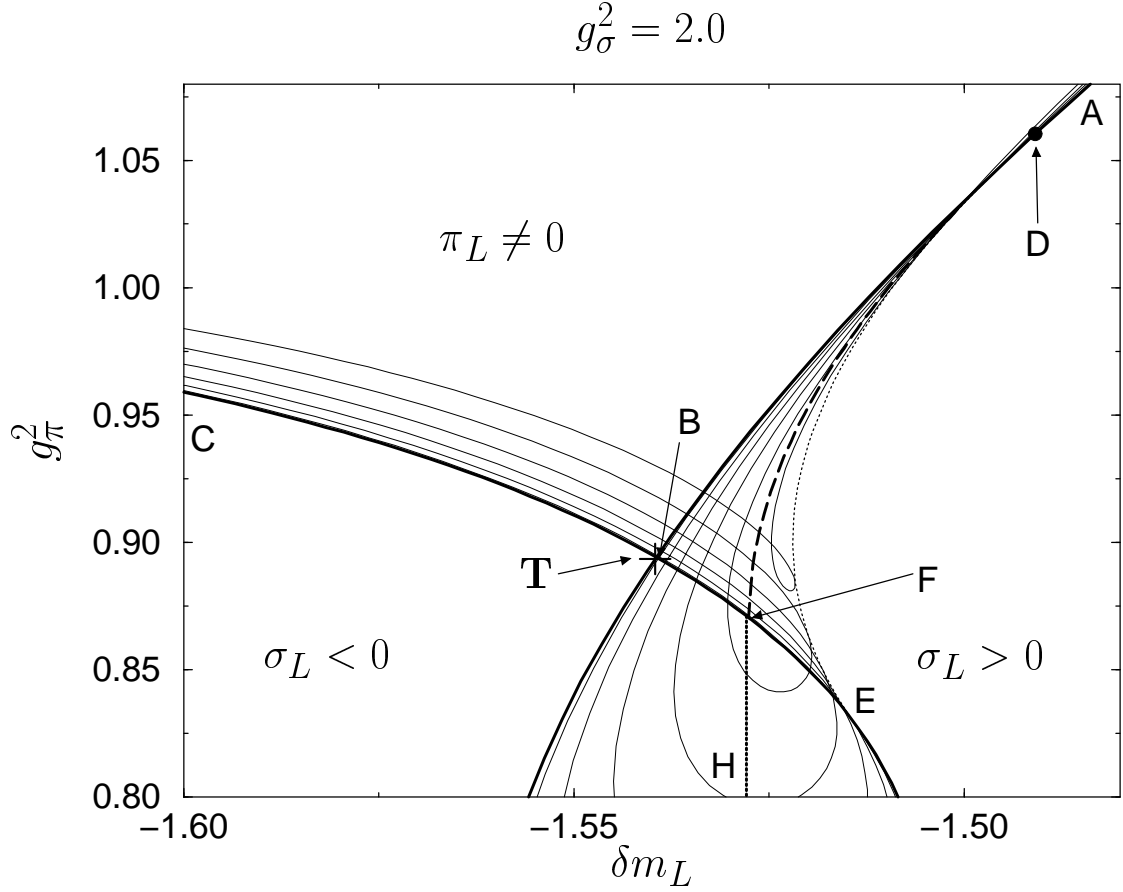} \par}

\vspace*{2cm}
\caption{Magnified view of the region near the tuned point 
${\bf T}(g_\sigma^2)$ for $g_\sigma^2=2.0$. 
Thick curve represents the parity phase boundary while thin lines are 
contour curves of fixed \protect\( \pi_L \protect \) values 
(\protect\( \pi_L =\, 0.02,\, 0.04,\, 0.06,\, 0.08,\, 0.10,\, 
0.12\protect \)). 
True phase boundary consists of AD (second order), DF(first order) and 
FBC(second order). 
\label{fig:mu0PhaseDet}} 
\end{figure}

In Fig.~\ref{fig:mu0PhaseDet} we plot an expanded view of 
the region near the tuned point ${\bf T}(g_\sigma^2)$; we take 
$g_\sigma^2=2.0$ for this 
figure as the structure is enhanced and so easier to draw. 
The thick curve represents the parity phase boundary, while 
the set of thin curves are lines of constant $\pi_L$, as determined from 
(\ref{124}-\ref{125}), for $0\leq \pi_L\leq 0.12$ in steps of 0.02.  
A number of important features are revealed in this figure. 
First of all, the boundary curve crosses with itself at the point B  
(this point differs from the tuned point ${\bf T}(g_\sigma^2)$ 
marked by a cross, to be discussed below).   
To understand why the crossing takes place, 
we note that the right-hand side of (\ref{121}) behaves as 
$ (\sigma_L\ln \sigma_L^2)/2\pi+2C_1$ near $\sigma_L=0$.
Hence
(\ref{121}) regarded as an equation of $\sigma_L$ for a given $\delta m_L$ 
has three solutions in the neighborhood of $\delta m_L=-2C_1 g_\sigma^2$, 
leading to the self-crossing behavior. 
Since the existence of three solutions is valid for arbitrarily small 
values of $g_\sigma^2$, the crossing 
persists down to $g_\sigma^2=0$.  Quite clearly there is no guarantee 
that the boundary curve represents the true phase boundary below 
the crossing point.

Another important feature is that the lines of constant $\pi_L$ spill 
beyond the boundary curve to the right of the crossing point.  
In the triangular region DBE formed by the two segments of the boundary curve 
DB-BE and the envelope ED 
of lines of constant $\pi_L$, the saddle point equations 
(\ref{124}-\ref{125}) have  
three solutions, one with $\pi_L=0$ and two more with $\pi_L\ne 0$.
Which of the solutions represent the true minimum can only be determined 
by examining the value of the effective potential.  
This is a typical situation where a first-order phase transition occurs.
   
A detailed numerical investigation of the effective potential shows that 
the true phase boundary runs as follows. 
Starting from right, the true phase boundary coincides with 
the parity phase boundary, and hence being of second order, 
from the point A to the point D where the envelope of lines of 
constant $\pi_L$ starts out.  Beyond this point, the phase transition becomes 
of first order, and runs along the dashed curve from the point D to 
the point F which is located on the parity phase boundary curve below 
the crossing point B.  At the point F, 
the phase transition returns to second order, and runs along the boundary 
curve from the point F through the crossing point B toward the point C.  
Altogether the parity-broken phase occupies the region above the line 
ADFBC.

Below this line the ground state of the model can be examined by 
the effective potential with $\pi_L=0$.  The potential no longer depends on 
$g_\pi^2$, and as a function of $\sigma_L$ has a double well structure.  
As one varies $\delta m_L$ below the point F from right to left, 
the true minimum switches from a positive to a negative value of $\sigma_L$ 
discontinuously.  This leads to a first-order phase transition along 
the line FH reaching down to $g_\pi^2=0$.  We shall call this 
line as the {\it $\sigma$ phase boundary}. 
Qualitatively, one may describe the behavior of the system near and below the 
point F as ``almost'' exhibiting spontaneous breakdown 
of chiral symmetry.  

We emphasize that the intricate structure involving first-order transitions 
described above originates from 
the ${\cal O}(a^3)$ term of the effective potential $V_L$ ({\it i.e.,} 
${\cal O}(a)$ in physical units) which contains terms cubic in $\sigma_L$.  
This contrasts with the leading ${\cal O}(a^2)$ terms in (\ref{60}) which 
gives rise to only simple second-order transitions 
on the $(\delta m_L,g_\pi^2)$ plane.
Indeed, including the ${\cal O}(a^3)$ contributions given by 
\begin{eqnarray}
\delta V_{L}=
-\frac{8}{3}C_{3}\sigma _{L}^{3}+
2\left[C_{1}'-\frac{1}{8\pi }\ln (\sigma _{L}^{2}+\pi ^{2}_{L})\right]
\sigma _{L}(\sigma _{L}^{2}+\pi ^{2}_{L}) \label{eq:dV3rd}
\end{eqnarray}
with 
\begin{eqnarray}
C_3=0.0647\cdots,\qquad C_{1}=0.0904\cdots , 
\end{eqnarray}
to the effective potential (\ref{60}), one quantitatively reproduces 
the features of the phase diagram shown in Fig.~\ref{fig:mu0PhaseDet}.  

\begin{figure}
\epsfxsize=0.9\textwidth
{\centering \epsfbox{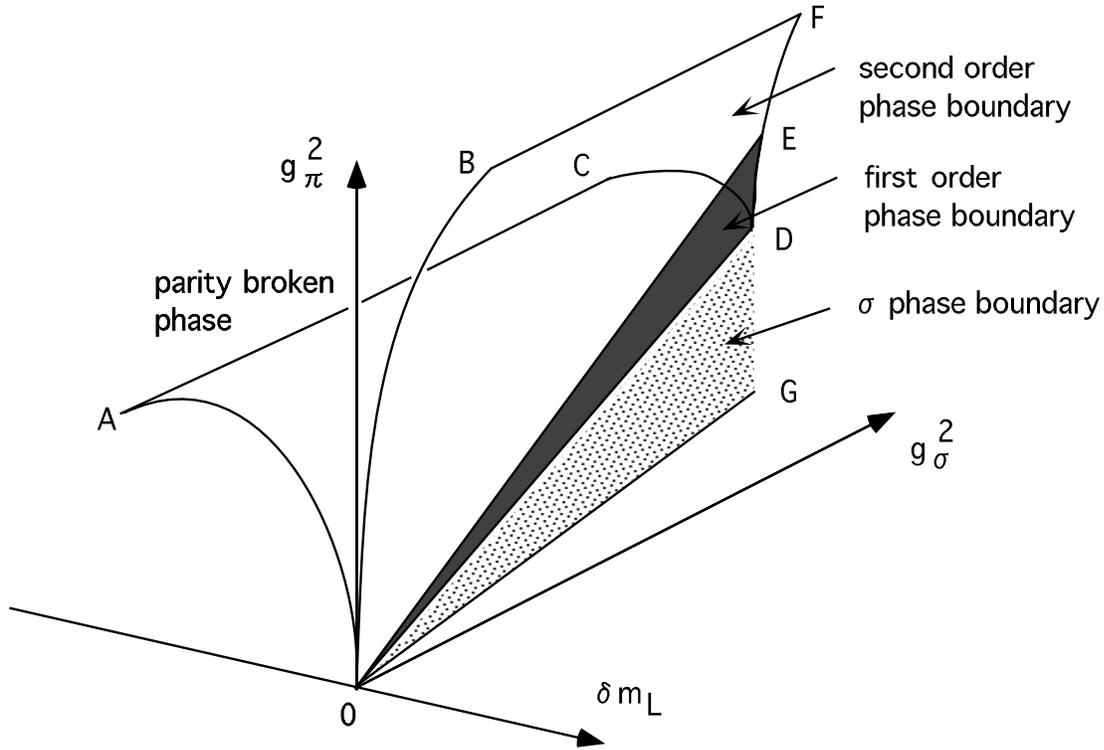} \par}

\vspace*{2cm}
\caption{Schematic phase diagram for $N_T=\infty$ and $\mu_L=0$. 
The phase boundary of the parity-broken phase AOB-CDEF
forms a blade whose edge touches the point 
O \protect\( (g^{2}_{\sigma },g^{2}_{\pi },\delta m_{L})=(0,0,0)\protect \). 
The first order part EOD shrinks to a point at $g_\sigma^2=0$. 
Parity-symmetric phase is divided into two phases 
with \protect\( \sigma_L <0\protect \)
and \protect\( \sigma_L >0\protect \) by a first-order $\sigma$ phase 
boundary DOG  which vertically drops from the edge OD. 
\label{fig:f4}}
\end{figure}

The phase structure on the $(\delta m_L, g_\pi^2)$ plane 
we have discussed for $g_\sigma^2=2$ remains the same 
toward $g_\sigma^2\to 0$ except that the structure as a whole moves down 
toward $g_\pi^2=0$ and becomes narrower horizontally.  
In the weak-coupling limit $g_\sigma^2=0$, the boundary line degenerates to 
$1/g_\pi^2=F(\delta m_L, 0)$ which does not have a self-crossing. 
The first-order line DF disappears, and the entire phase boundary 
becomes of second order. 
We thus find the phase structure in the three-dimensional parameter space  
$(g_\sigma^2, g_\pi^2, \delta m_L)$ as 
schematically drawn in Fig.~\ref{fig:f4}.   

Let us now discuss the continuum limit, in particular, 
how the phase structure relates with the tuning of parameters 
specified by (\ref{110}-\ref{91}).  
The tuned point ${\bf T}(g_\sigma^2)$ without ${\cal O}(a)$ terms
(cross in Fig.~\ref{fig:mu0PhaseDet}) 
is located slightly to the left and below 
the point B in the parity-symmetric phase.  When one takes the coupling 
$g_\sigma^2$ to zero, the relative location of the tuned point and the 
$\sigma$ phase boundary remains the same.  
Therefore taking the continuum limit exactly 
along the tuned point ${\bf T}(g_\sigma^2)$ leads to a negative value of 
$\sigma$.  This, however, does not mean that a continuum limit with a 
positive value of $\sigma$ is not possible.  

As we have already pointed out, the 
structure around the point F stems from the ${\cal O}(a^3)$ terms of the 
effective potential, and hence its size shrinks as a power of $a$.  
One can estimate the effect of adding (\ref{eq:dV3rd}) to (\ref{60}) 
by an order counting.
Since $\delta m_L$ is the coefficient of the linear term $\sigma_L$ 
in (\ref{60}), the ${\cal O}(a^3)$ correction (\ref{eq:dV3rd}) causes a shift 
of form   
\begin{equation}
\delta m_L \rightarrow \delta m_L + {\cal O}(\sigma_L^2) 
\sim \delta m_L + {\cal O}(a^2) \,.
\end{equation}
Hence the structure has a size of ${\cal O}(a^2)$ in $\delta m_L$. 
By a similar argument for $g_\pi^2$, we conclude a size of 
${\cal O}(a)$ for $g_\pi^2$. 
The point of tuning, on the other hand, has a degree of 
freedom of ${\cal O}(a^2)$ for $\delta m_L$ and 
${\cal O}(a)$ for $g_\pi^2$ . 
Therefore one may shift the point of tuning within a rectangular region 
of size ${\cal O}(a^2)$ times ${\cal O}(a)$ around the point 
${\bf T}(g_\sigma^2)$, including a shift to the right of the point F, 
in which case the continuum value of $\sigma$ becomes positive. 

\begin{figure}
\epsfxsize=0.9\textwidth
{\centering \epsfbox{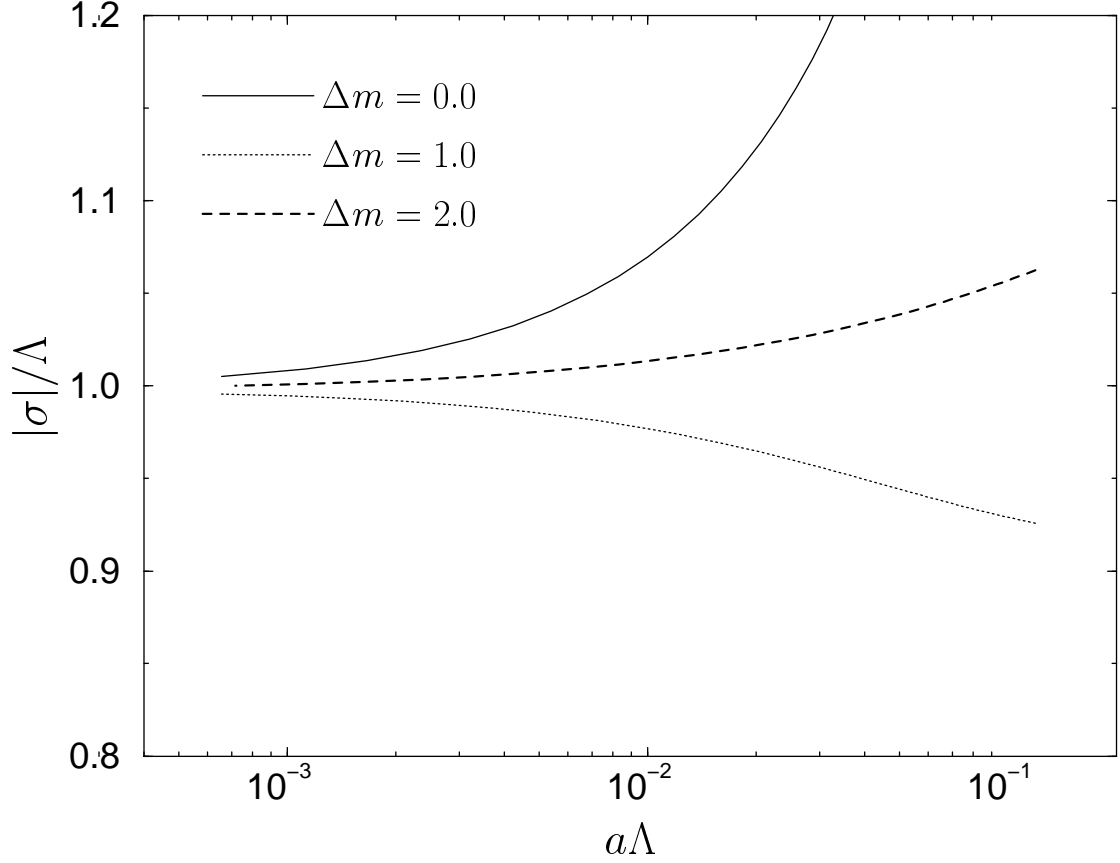} \par}

\vspace*{2cm}
\caption{Order parameter \protect\( \sigma \protect \) in units of 
$\Lambda$ as a function of lattice spacing 
\protect\( a\Lambda \equiv c\exp (-\pi /g^{2}_{\sigma })\protect \) 
calculated for three choices of continuum extrapolation specified by 
$\Delta m$ for \protect\( \Delta g_\pi=0 \protect \).
Sign of $\sigma$ is negative for $\Delta m=0$. 
\label{fig:SGscale}}
\end{figure}

In order to demonstrate that the correct continuum limit is obtained within 
the freedom of the choice of the parameters described above, 
we examine the scaling of two physical observables 
$\sigma/\Lambda$ and $(M_\pi/\Lambda)^2$ along the lines given by 
\begin{eqnarray}
\delta m_L &=& -2C_1 g_\sigma^2 + \Delta m (a \Lambda)^2,
\label{eq:mLarb}\\
{1\over g_\pi^2} &=& {1\over g_\sigma^2} + 4 C_2 + \Delta g_\pi a \Lambda
\label{eq:gpiarb}\,,
\end{eqnarray}
where $\Delta m$ and $\Delta g_\pi$ are ${\cal O}(1)$ constants.
We show the solution of the saddle point equation (\ref{124}) for 
$\sigma/\Lambda=\sigma_L/(a\Lambda)$ for 
$\Delta m= 0,1,2$ and $\Delta g_\pi=0$ 
in Fig.~\ref{fig:SGscale}
as a function of $a\Lambda=c \exp(-\pi/g_\sigma^2)$ where $\Lambda$ is 
defined in (\ref{65}). 
One sees that $\sigma/\Lambda$ converges to the correct continuum result
$\sigma/\Lambda=1$ whether the limit is taken on the left ($\Delta m=0$) or 
the right ($\Delta m= 1,2$) of the point F of the phase boundary.  

\begin{figure}
\epsfxsize=0.9\textwidth
{\centering \epsfbox{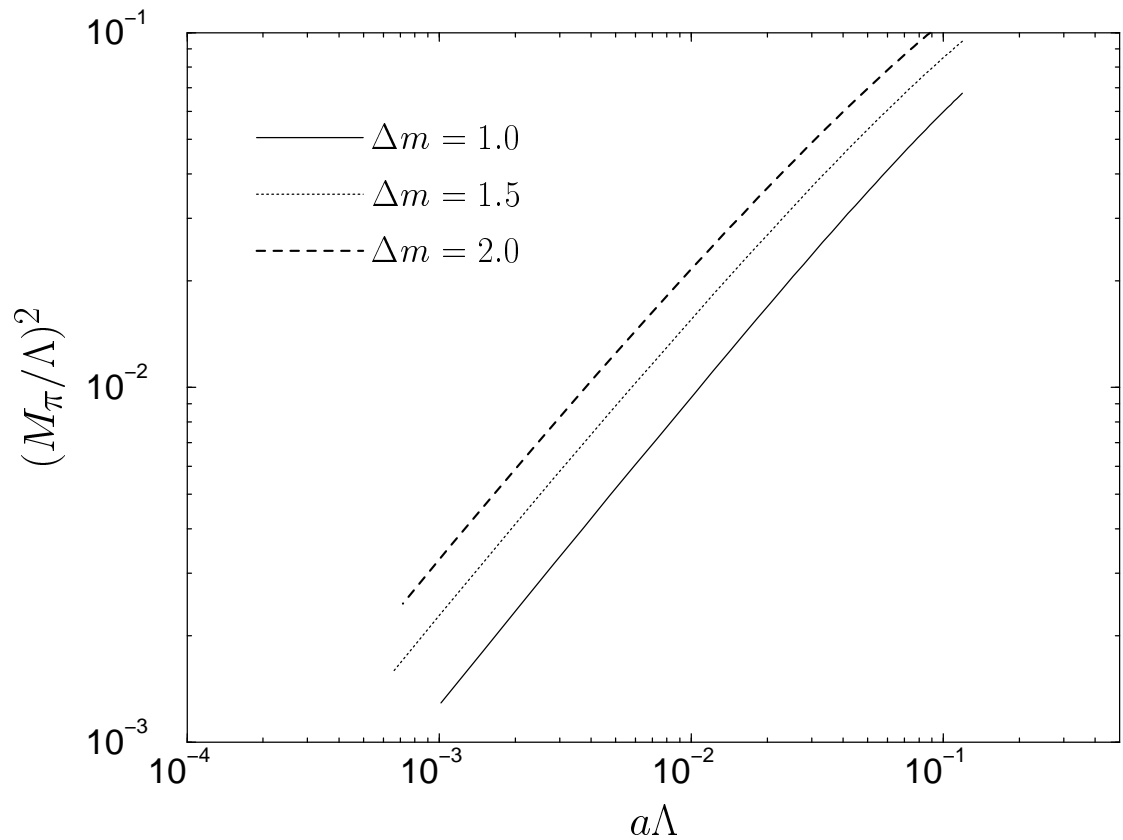} \par}

\vspace*{2cm}
\caption{Pion mass squared as a function of 
lattice spacing \protect\( a\Lambda\protect \) under continuum extrapolation.
See text for tuning of couplings employed. 
\label{fig:PiMass}}
\end{figure}

The pion mass $M_\pi$ is more interesting since the 
mechanism\cite{Aoki1,AUU} which ensures $M_\pi=0$ even for finite $a$ no 
longer works along the segment DF in Fig.~\ref{fig:mu0PhaseDet} 
because of the first-order nature of 
the phase transition in this part of the phase boundary. 
We select $\Delta m = 1, 1.5, 2.0$ and tune $\Delta g_\pi$ so that the point 
$(\delta m_L, g_\pi^2)$ given by (\ref{eq:mLarb}-\ref{eq:gpiarb}) is 
placed just under the first-order line DF in the parity-symmetric phase.
Solving the saddle point equations (\ref{124}-\ref{125}) for $\sigma_L$ 
and substituting the result into (\ref{150}), we obtain the curve for 
$(M_\pi/\Lambda)^2$
plotted in Fig.~\ref{fig:PiMass} as a function of $a\Lambda$.  
The squared pion mass 
vanishes as a power of $a$ toward the continuum limit, demonstrating that 
the first-order nature of the parity-breaking phase
transition along DF does not cause problems. 

So far we have concentrated on the phase structure near $\delta m_L=0$.  
The phase structure around $\delta m_L=-4$ is obtained by a reflection 
$\delta m_L\to -4-\delta m_L$. The phase structure near $\delta m_L=-2$
is different.  It is symmetric across the 
$\delta m_L=-2$ plane, and the parity phase boundary turned out to be  
always of second order.  In addition 
a $\sigma$ phase boundary of first-order transition across which 
the sign of $(\sigma_L+2)$ flips exists at $\delta m_L=-2$, 
similar to the line FH in Fig.~\ref{fig:mu0PhaseDet}. 

In the following we concentrate on the phase structure 
around  $\delta m_L=0$ relevant for the usual continuum limit. 

\section{Phase Diagram at $T\neq 0$ and $\mu=0$}
\label{sec:four}

Our analysis of the phase diagram for the case of finite temperature 
and zero chemical potential proceeds in essentially the same 
way as for the zero-temperature case; choosing a temporal lattice size $N_T$ 
and a value of $g_\sigma^2$, we analyze the solution of the saddle point 
equations and the value of the effective potential given by 
(\ref{170}-\ref{230}) with $\mu_L=0$ as 
a function of $g_\pi^2$ and $\delta m_L$.  

\begin{figure}
\epsfxsize=0.9\textwidth
{\centering \epsfbox{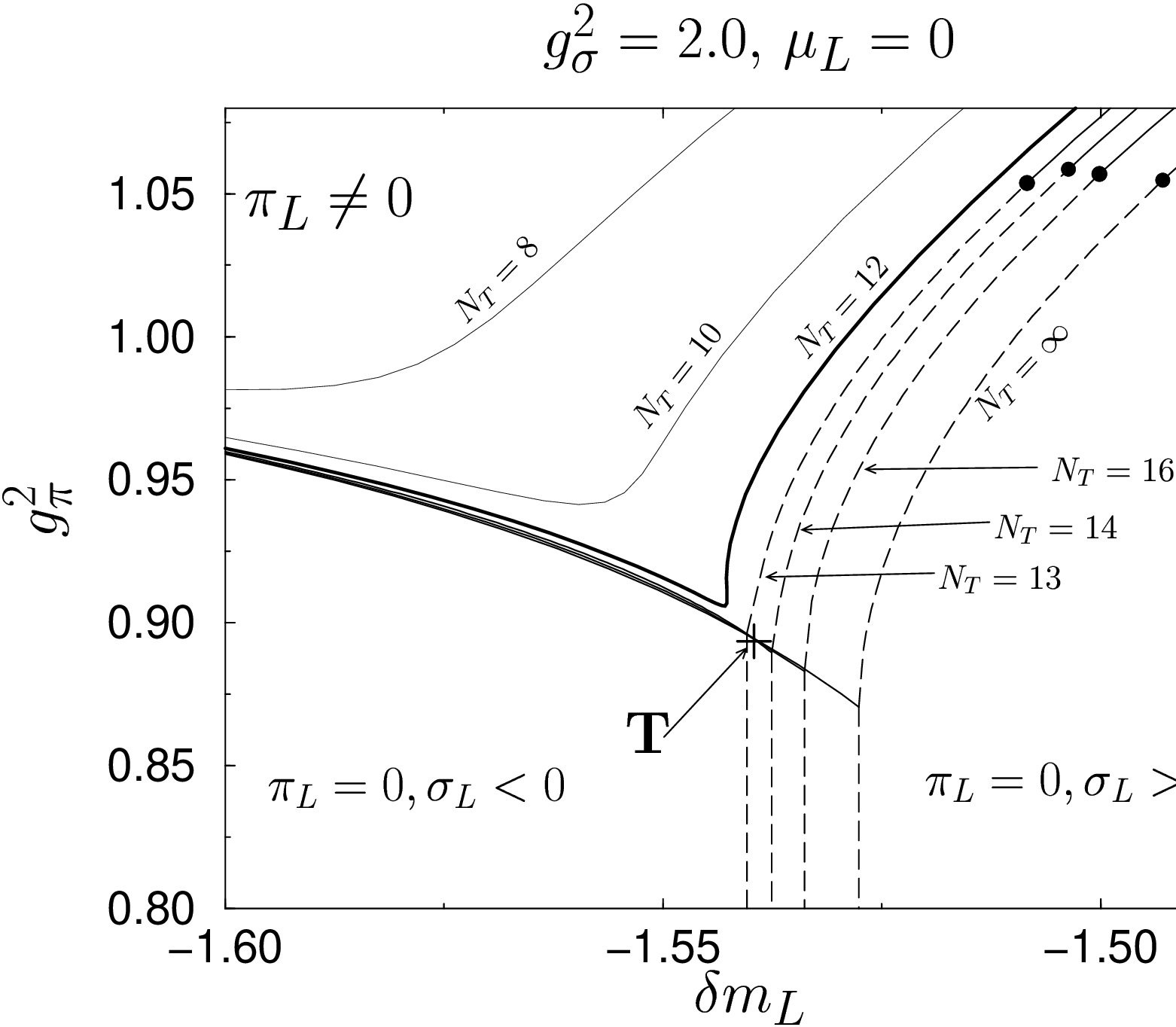} \par}

\vspace*{2cm}
\caption[]{Temporal lattice size dependence of phase
 structure plotted for $N_{T}=\infty$, 16, 14, 13, 12, 10, 8.
Phase transition is of first order along dashed lines, while it is of second 
order along solid lines. 
\label{fig:PhaseNt}}
\end{figure}

In Fig.~\ref{fig:PhaseNt} we show the result of 
such an analysis at $g_\sigma^2=2$ 
for a set of temporal lattice size $N_T$. 
For large $N_T$, the topological structure of the phase boundary remains the 
same as at $N_T=\infty$ corresponding to zero temperature.  
In particular the left-half of the phase boundary hardly changes.  
On the other hand, the right-half of the boundary with the first-order segment
moves toward upper left. The lower end point of the first-order 
segment slides up along the second-order part of the boundary 
for $N_T=\infty$.  
The vertical $\sigma$ phase boundary also moves toward left with the end 
point. 

This structure changes at a threshold value 
of $N_T$, which equals $N_T=13$ for $g_\sigma^2=2$ chosen for 
Fig.~\ref{fig:PhaseNt}, 
when the first-order segment of the boundary disappears.  
The entire phase boundary becomes a smooth line of second-order 
transition, along which $\sigma_L$
continuously varies from positive to negative values as one traverses the 
boundary curve from right to left.  As may be inferred from this, 
the vertical $\sigma$ phase boundary also disappears at the threshold 
temporal lattice size.  

\begin{figure}
\epsfxsize=0.9\textwidth
{\centering \epsfbox{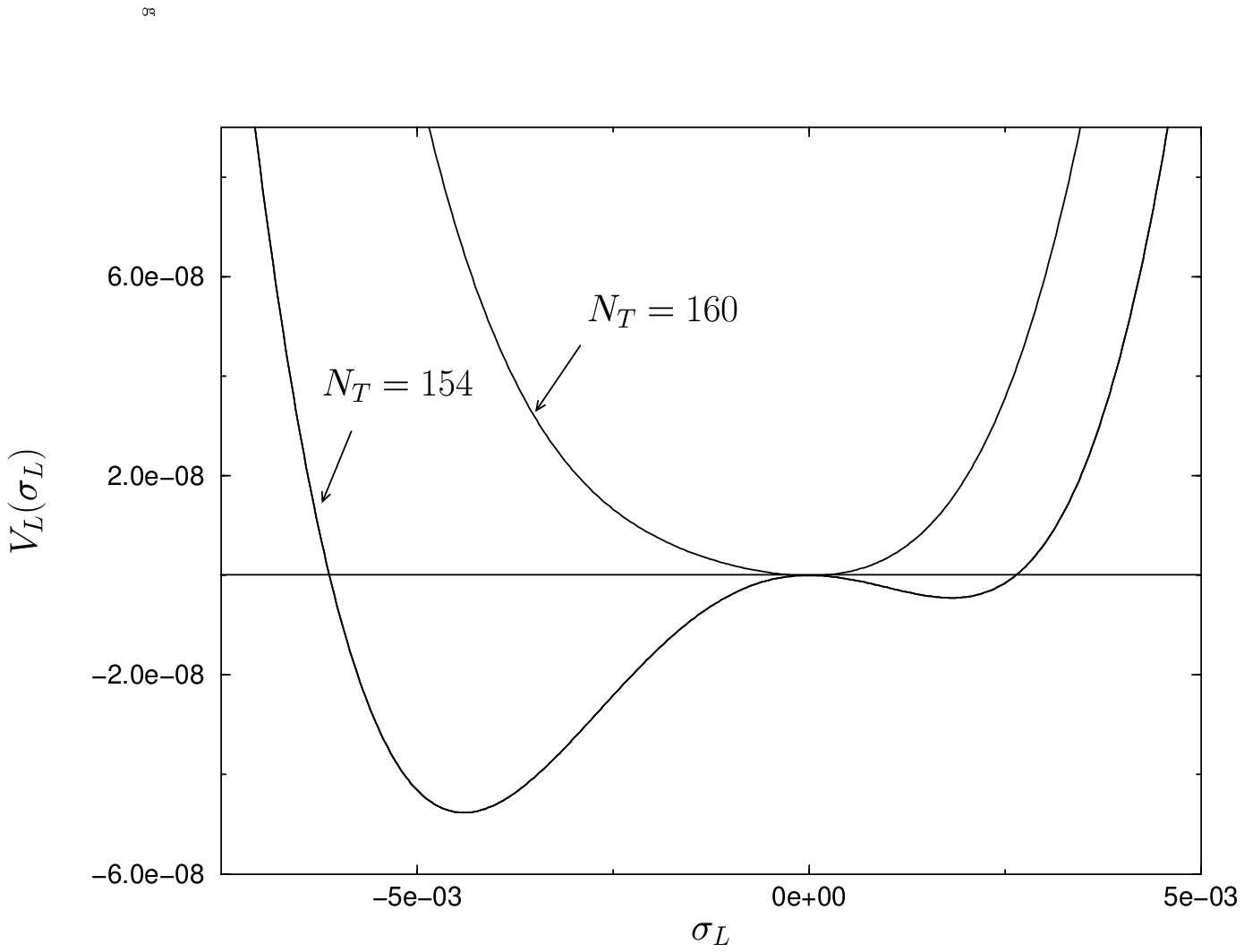} \par}

\vspace*{2cm}
\caption{Effective potential for  $g_\sigma^2=0.8$ for temporal lattice 
size \protect\( N_{T}=160\protect \) and 
\protect\( N_{T}=154\protect \). $\delta m_L$ is tuned according 
to ${\bf T}(g_\sigma^2)$.
\label{fig:f7}}
\end{figure}

To the extent that the first-order $\sigma$ 
boundary is interpreted as indicative of spontaneous breakdown of 
chiral symmetry, 
its disappearance may be regarded as signifying restoration of chiral 
symmetry.  Indeed the value of $\sigma_L$ becomes small when the $\sigma$
phase boundary disappears. 
We illustrate this point in Fig.~\ref{fig:f7} which shows how the effective 
potential changes from a double-well to a single-well structure 
across the threshold value of $N_T$ at the tuned point 
${\bf T}(g_\sigma^2)$ for $g_\sigma^2=0.8$. 

\begin{figure}
\epsfxsize=0.9\textwidth
{\centering \epsfbox{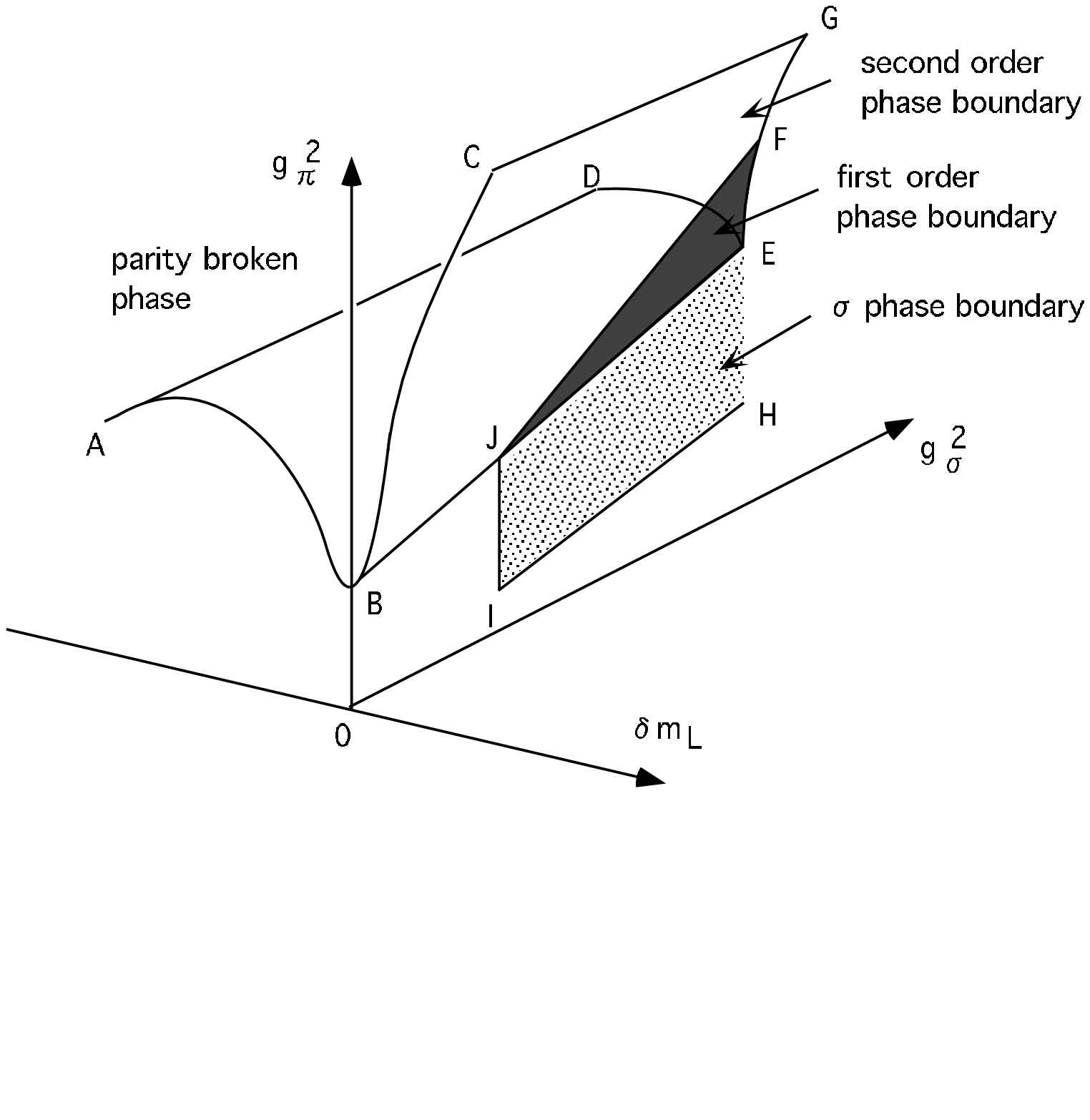} \par}

\vspace*{2cm}
\caption{Schematic phase diagram for fixed and finite 
\protect\( N_{T}\protect \). 
For small \protect\(g^{2}_{\sigma }\protect \), 
first-order parity-breaking boundary (FJE) and \protect\( 
\sigma \protect \) phase boundary (EJIH) disappear. 
\label{fig:f6}}
\end{figure}

The structural change found as a function of $N_T$ for a fixed $g_\sigma^2$ 
can be translated into that for a varying $g_\sigma^2$ with a fixed 
temporal size $N_T$ : as one lowers $g_\sigma^2$, there is a threshold value 
$g_\sigma^2=g_{\sigma c}^2(N_T)$ 
below which the first-order segment of the parity-broken
phase boundary and the vertical first-order $\sigma$ boundary cease to exist. 
This leads to the three-dimensional phase structure for a fixed and finite 
$N_T$ schematically drawn in Fig.~\ref{fig:f6}. 
Contrary to the $N_T=\infty$ case, 
the boundary curve at $g_\sigma^2=0$ does not reach the value 
$g_\pi^2=0$ since   
it is given by $1/g_\pi^2=F(\delta m_L, 0)$ where the function 
$F(\delta m_L, 0)$ computed for a finite $N_T$ is finite and positive at its 
maximum at $\delta m_L=0$.

Let us consider the continuum limit of the model at finite temperature. 
This limit is defined by simultaneously taking $N_T\to\infty$ and 
$g_\sigma^2\to 0$ such that 
\begin{equation}
\frac{T}{\Lambda}=\frac{\exp (\pi/g_\sigma^2)}{cN_T}
\end{equation}
is kept fixed.  The coupling $g_\pi^2$ and the mass 
parameter $\delta m_L$ have to be tuned according to (\ref{110}-\ref{91}) 
to restore chiral symmetry. 
An alternative way to take the limit, which is closer to 
the method practiced in numerical simulations, is to vary $g_\sigma^2$ for a 
fixed $N_T$. For each $N_T$ this yields physical observables as a function of 
$T/\Lambda$, and the continuum limit is obtained as a limit of this function 
as $N_T\to\infty$. 
We follow the latter approach in the following analysis. 

A qualitative picture of how finite-temperature transition appears in this
approach is as follows.  The continuum limit is taken along the line 
satisfying (\ref{110}-\ref{91}), which converges to 
$(g_\sigma^2, g_\pi^2, \delta m_L)=(0,0,0)$.  
Above the threshold value $g_\sigma^2=g_{\sigma c}^2(N_T)$, 
this line runs along either 
side of the lower edge of the parity-broken phase where physical 
quantities take values similar to those at zero temperature. 
In particular $\sigma_L$ has a non-zero value of positive or negative sign 
depending on whether the line is on the right or left of the vertical $\sigma$ 
phase boundary. 
However, as $g_\sigma^2$ is lowered below the threshold value, 
the parity-breaking phase boundary moves up toward larger $g_\pi^2$, and the 
$\sigma$ phase boundary disappears.  Once the line enters into this region, 
the value of $\sigma_L$ becomes small. In other words, 
for each fixed temporal size, 
the threshold value of $g_\sigma^2$ marks the region of 
finite-temperature transition. 

\begin{figure}
\epsfxsize=0.9\textwidth
{\centering \epsfbox{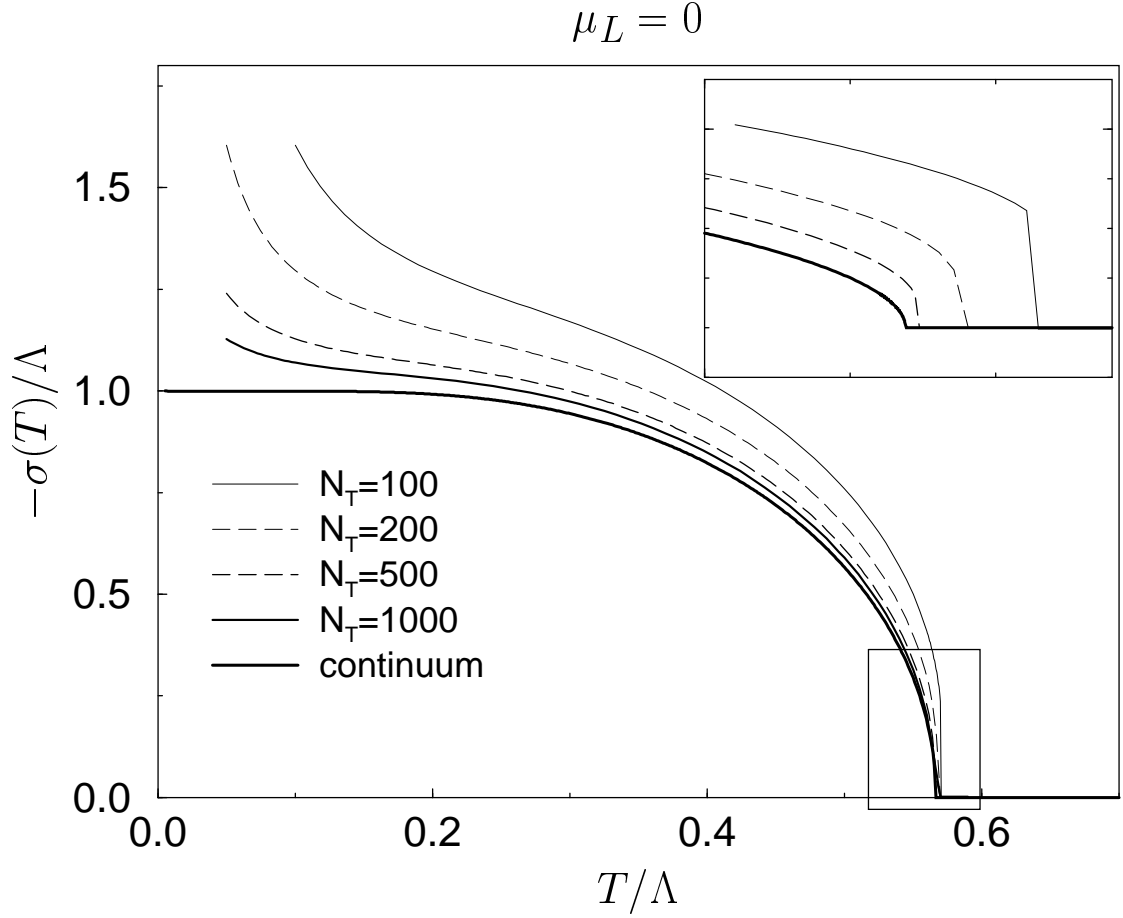} \par}

\vspace*{2cm}
\caption{Order parameter \protect\( \sigma \protect \) 
as a function of temperature \protect\( T\protect \) for 
\protect\( \mu =0\protect \) evaluated for several temporal lattice size 
along the point ${\bf T}(g_\sigma^2)$ without ${\cal O}(a)$ corrections.
Inset shows an expanded view of the critical region. 
\label{fig:sgNT1loop}}
\end{figure}

We illustrate the description above in Fig.~\ref{fig:sgNT1loop} where 
we plot $\sigma/\Lambda$ as a function of $T/\Lambda$ 
calculated along the line ${\bf T}(g_\sigma^2)$ without ${\cal O}(a)$ 
corrections for several values of $N_T$, together with the curve 
in the continuum limit.  We observe that the curves for finite $N_T$ 
smoothly approaches the continuum result drawn by a thick line as $N_T$
increases toward $\infty$.  

Several remarks are in order with this figure: 
(i) The negative sign of $\sigma$ is due to the fact, already 
remarked in Sec.~\ref{sec:three}, 
that the continuum limit is taken on the left side
of the $\sigma$ phase boundary. 
(ii) In the region close to the continuum critical point 
$T/\Lambda\approx T_c/\Lambda=\exp (\gamma_E/\pi)=0.56694\cdots$, 
$\sigma(T)$ has a small discontinuity, as shown in the inset of 
Fig.~\ref{fig:sgNT1loop}, which reduce with increasing $N_T$.  
This is due to the fact that the tuned point ${\bf T}(g_\sigma^2)$ goes 
through the vertical $\sigma$ phase boundary as $g_\sigma^2$ is decreased 
just before the threshold value is reached 
(see the relative location of the boundary of $N_T=13$  
and ${\bf T}(g_\sigma^2)$ in Fig.~\ref{fig:PhaseNt}). 
(iii) The value of $\sigma (T)$ after the discontinuity is very small but
non-zero for a finite $N_T$.  It exactly vanishes for $T\geq T_c$ only 
in the limit of $N_T\to\infty$. 

Another possible choice for the continuum limit is 
to run along the bottom of the valley of the parity-breaking phase boundary 
for each value of $N_T$ (line EJB in Fig.~\ref{fig:f6}).  
The limiting point B of the line at $g_\sigma^2= 0$ has a non-zero value 
of $g_\pi^2$.  This point, however, reaches the correct continuum limit 
$(g_\sigma^2, g_\pi^2, \delta m_L)=(0,0,0)$ as $N_T\to\infty$. 
Above the threshold value of $g_\sigma^2$ (point J) where 
the $\sigma$ phase boundary exist, $\sigma(T)$ has two solutions of opposite
sign($\sigma^+_L,\sigma^-_L$). At the bottom of the valley,
these two solution are the balancing minima of the effective 
potential, {\it i.e.},
\begin{equation}
V_{L}(\sigma^+_L)=V_{L}(\sigma^-_L)\,.
\end{equation}
At the threshold value of $g_\sigma^2$, $\sigma^+_L$ and $\sigma^-_L$ 
merge through a second-order singularity corresponding to the
second-order phase transition (line JI) marking the end of the 
first-order $\sigma$ phase boundary.

\begin{figure}
\epsfxsize=0.9\textwidth
{\centering \epsfbox{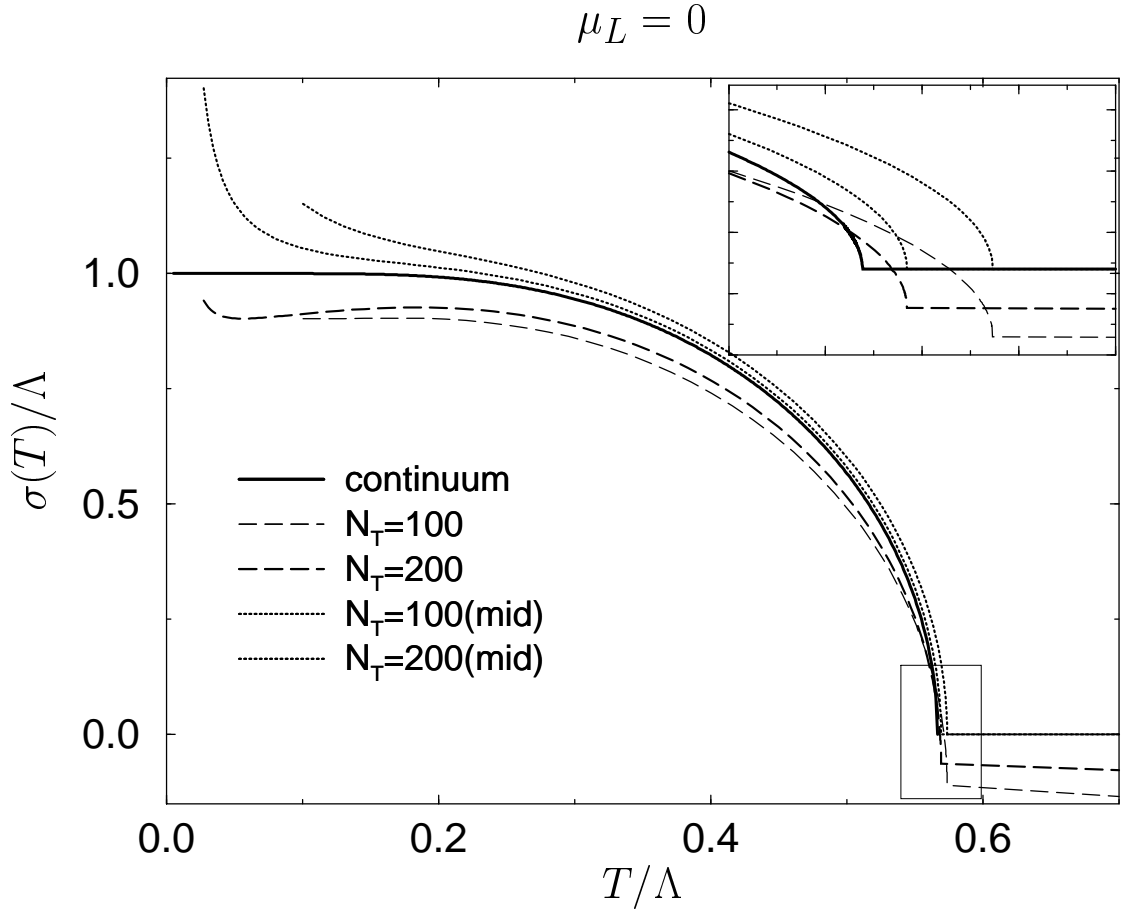} \par}

\vspace*{2cm}
\caption{Positive solution $\sigma^+$ as a function of $T$ for $\mu=0$.  
Couplings are tuned along the edge of the parity-breaking phase boundary 
(line EJB in Fig.~\ref{fig:f6}) for each value of $N_T$ .
The combination $(\sigma^+-\sigma^-)/2$ is also plotted 
(see curves specified as ``(mid.)''). 
Inset shows an expanded view of the critical region.
\label{fig:sgNtBP}}
\end{figure}

\begin{figure}
\epsfxsize=0.9\textwidth
{\centering \epsfbox{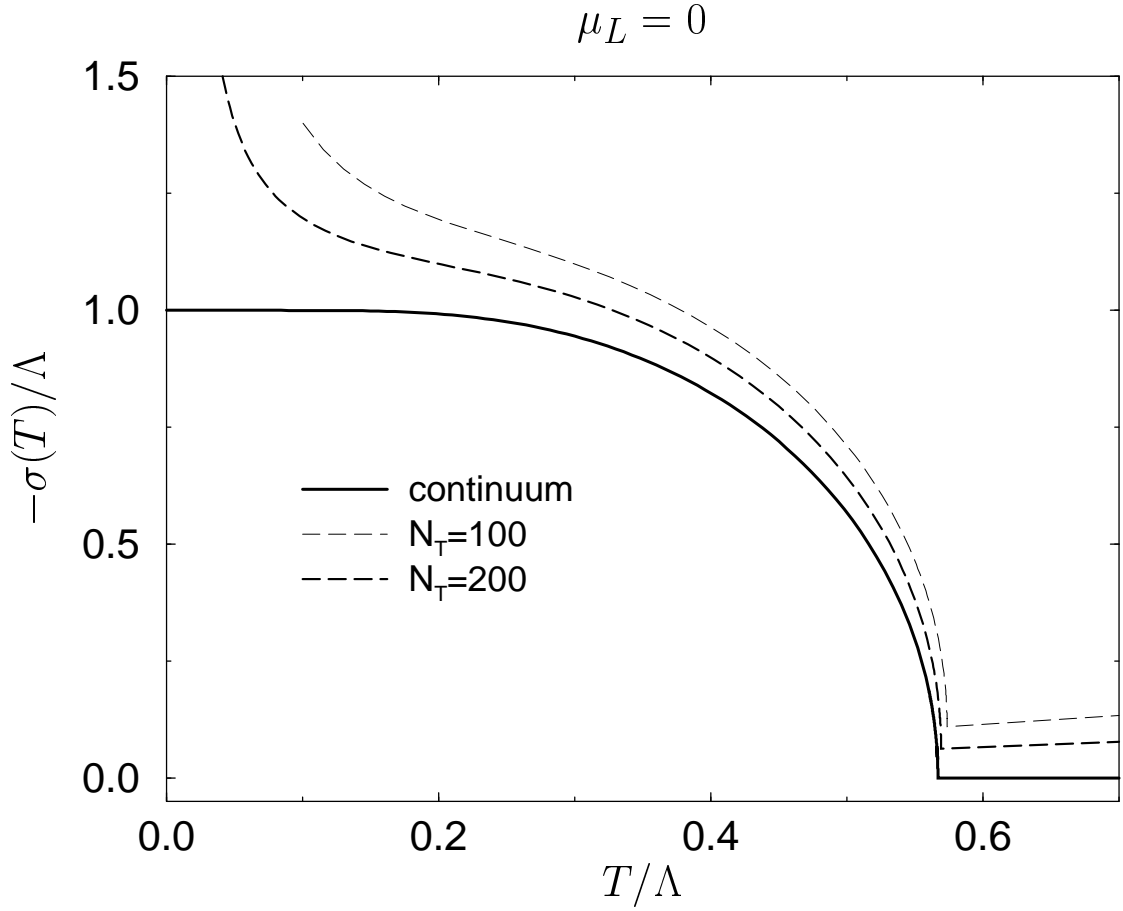} \par}

\vspace*{2cm}
\caption{Same as Fig.~\ref{fig:sgNtBP} for the negative solution 
$\sigma^-$.
\label{fig:sgNtBN}}
\end{figure}

With this choice of the continuum limit, we obtain the behavior of 
$\sigma^+ (T)$ and $\sigma^-(T)$ shown by dashed lines 
in Fig.~\ref{fig:sgNtBP} and Fig.~\ref{fig:sgNtBN}, 
exhibiting convergence to the continuum result (thick line).  
A notable property is that the system exhibits a 
second-order phase transition already for finite values of $N_T$ as the line
passes through the point J in Fig.~\ref{fig:f6}. 
Also shown in  Fig.~\ref{fig:sgNtBP} by dotted lines is the difference of the 
two solutions divided by 2, $(\sigma^+_L-\sigma^-_L)/2$, 
which shows a faster convergence to the continuum result. 

\section{Phase Diagram for $\mu\neq0$}
\label{sec:five}

\subsection{The Case of $\mu\neq0$ and $T=0$}
\label{subsec:fiveA}

\begin{figure}
\epsfxsize=0.9\textwidth
{\centering \epsfbox{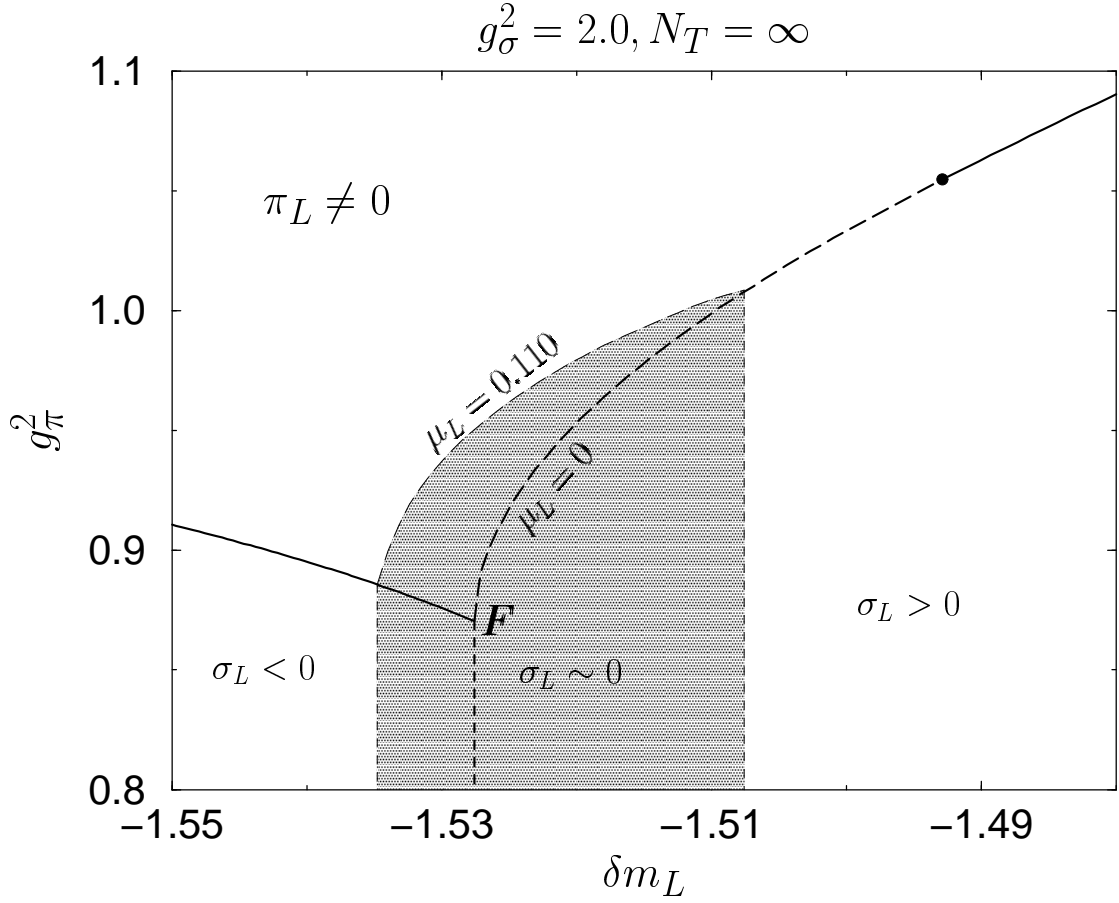} \par}

\vspace*{2cm}
\caption{Phase structure for $\mu_L=0.110$ and $N_T=\infty$ 
at $g_\sigma^2=2.0$. Hatched region is the new phase
with $\sigma_L\approx 0$. 
Phase transitions between this phase ($\pi_L=0, \sigma_L\approx 0)$
and other three phases (\protect\( \pi _{L} \neq 0,\, \sigma _{L} <0,\, 
 \sigma _{L} >0\protect \)) are all first order. 
\label{fig:muNtInf}}
\end{figure}

\begin{figure}
\epsfxsize=0.9\textwidth
{\centering \epsfbox{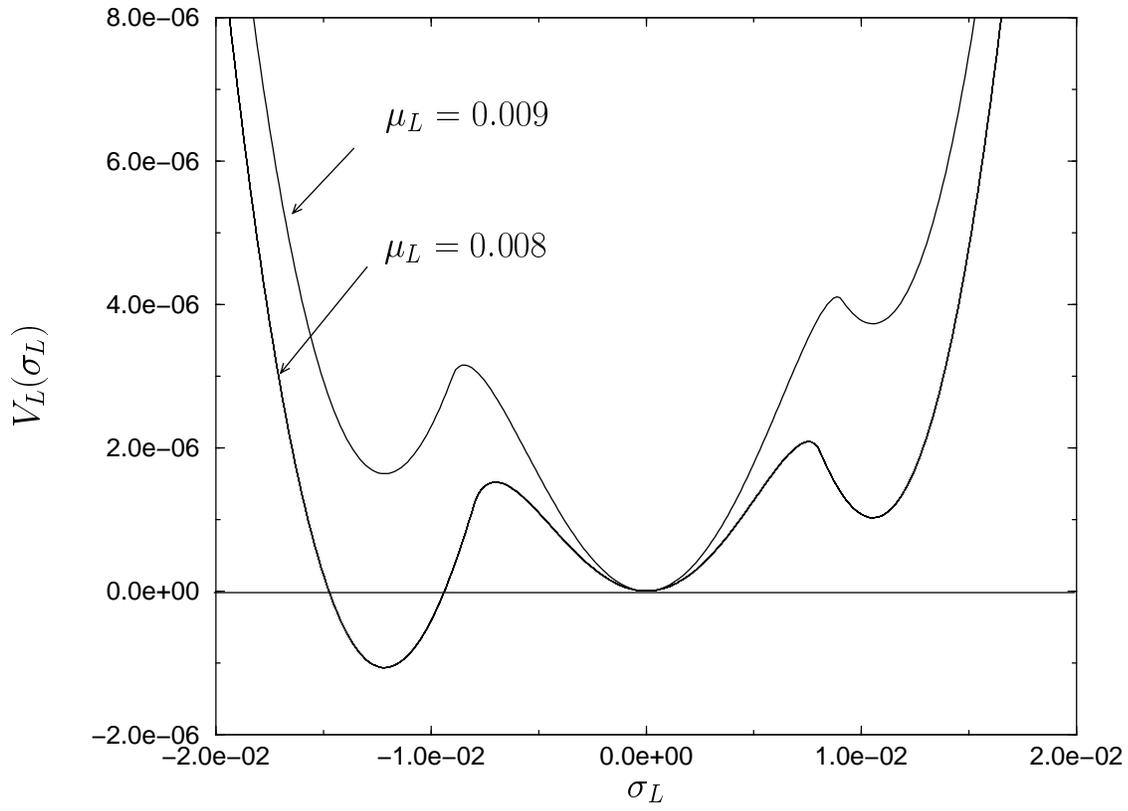} \par}

\vspace*{2cm}
\caption{Effective potential as a function of \protect\( \sigma _{L}
\protect \) for \protect\( g^{2}_{\sigma }=0.8 \protect \) and  
\protect \( N_{T}=\infty \protect \) at the tuned point 
${\bf T}(g_\sigma^2)$.
\label{fig:f10}}
\end{figure}

We now consider the phase diagram for non-zero chemical potential, 
starting with the case of zero temperature ($N_T=\infty$). 
In this case, from (\ref{170}-\ref{230}), the effective potential is given by  
\begin{eqnarray}
V_L=\frac{1}{2g_\sigma^2}(\sigma_L-\delta m_L)^2
         +\frac{1}{2g_\pi^2}\pi_L^2-\int_{-\pi}^{\pi}
        \!\!\frac{d\xi}{2\pi}\Big[\ln(\sigma_L+2-\cos\xi)+\max(E,\,\mu_L)
\Big] \label{250}.
\end{eqnarray}
As one expects from this expression,  as long as $\mu_L$ is small, 
the phase boundary on the $(\delta m_L, g_\pi^2)$ plane does not
move from that with $\mu_L=0$. 
However, as we show in Fig.~\ref{fig:muNtInf} for $g_\sigma^2=2$, 
when $\mu_L$ exceeds a threshold value ($\mu_{L c}\approx 0.09$ for 
Fig.~\ref{fig:muNtInf}), the vertical 
$\sigma$ phase boundary splits into two vertical first-order boundaries 
such that $\sigma_L$ between the two boundaries is near zero.  The end 
points of the two boundaries slide upwards in the opposite directions 
along the parity phase boundary for $N_T=\infty$, and the phase boundary 
connecting the two end points, which is of first-order,  
is convex and moves upward toward larger $g_\pi^2$. 
This behavior reflects the fact that the effective potential, which has 
a double well structure at $\mu_L\approx 0$ develops an additional minimum
close to $\sigma_L=0$ as $\mu_L$ increases beyond the threshold value.   
See Fig.~\ref{fig:f10}.  Which of the three minima represents the true 
minimum depends on the 
value of $\delta m_L$, which leads to the double first-order boundaries 
seen in Fig.~\ref{fig:muNtInf}. 

\begin{figure}
\epsfxsize=0.9\textwidth
{\centering \epsfbox{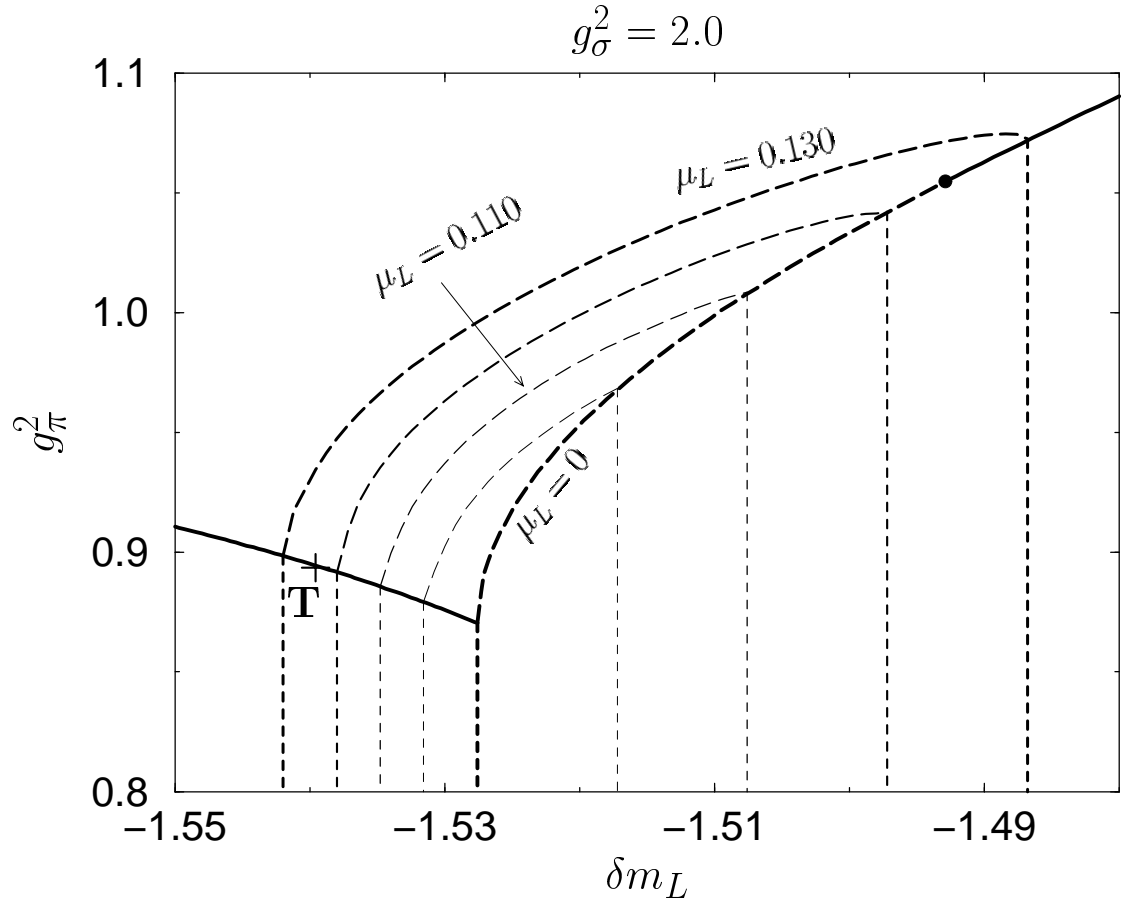} \par}

\vspace*{2cm}
\caption{Same as Fig.~\ref{fig:muNtInf} for \protect\( \mu _{L}=0,\,
 0.10,\, 0.11,\, 0.12,\, 0.13\protect \).
Symbol {\bf T} represent the tuned point ${\bf T}(g_\sigma^2)$. 
\label{fig:muNtInfB}}
\end{figure}

\begin{figure}
\epsfxsize=0.9\textwidth
{\centering \epsfbox{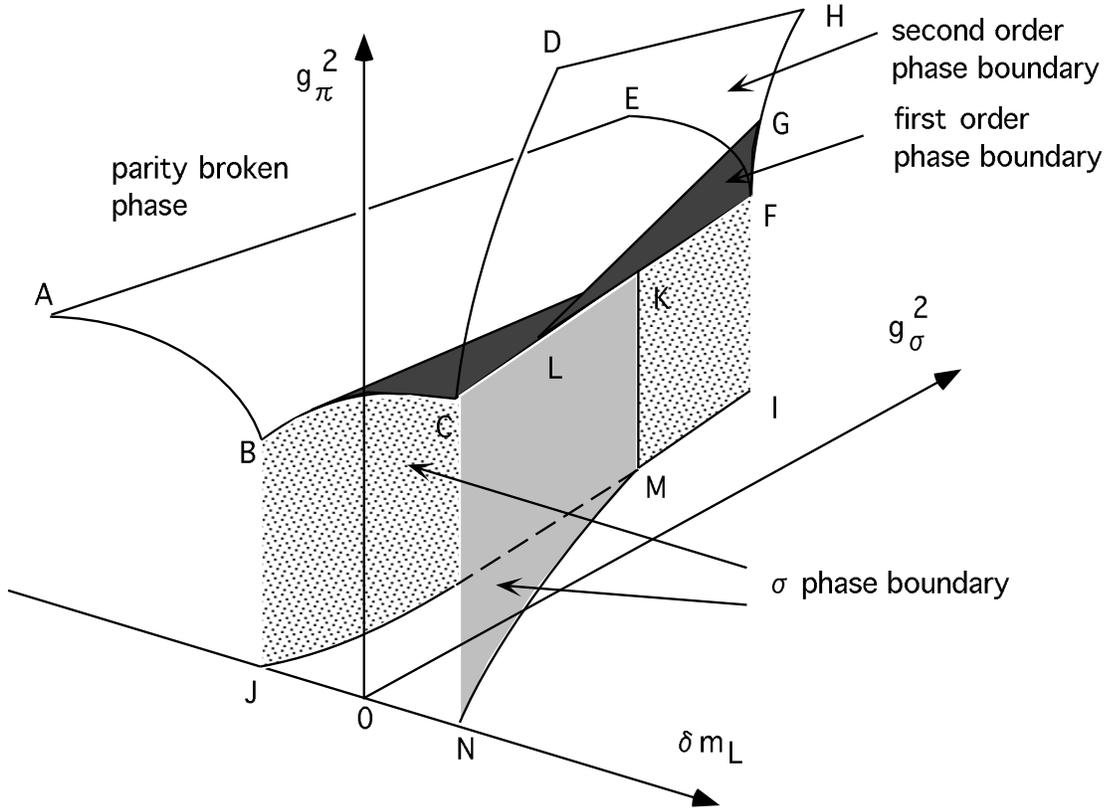} \par}

\vspace*{2cm}
\caption{ Sketch of phase structure for fixed and finite 
\protect \( \mu _{L} \protect \) for \protect \( N_{T}=\infty \protect \). 
Below a threshold value of $g_\sigma^2$ the  vertical $\sigma$ phase 
boundary splits into two, both being of first order, enclosing a region 
with $\sigma_L\approx 0$. \label{fig:f8}}
\end{figure}

An important point with the behavior described above is that the two 
vertical first-order boundaries move away from the edge point F of the 
parity phase boundary at $\mu_L=0$.  
Since the point (\ref{110}-\ref{91}) for taking the continuum 
limit is located near the point F, this means that the system, initially 
in one of the phases with a non-zero value of $\sigma_L$ at $\mu_L=0$ 
undergoes a first-order phase transition when the boundary moves over 
the point as $\mu_L$ increases.  After this takes place, the system is
in the middle phase with $\sigma_L\approx 0$, which qualitatively means
that chiral symmetry becomes restored.  
This, then, is the mechanism by which the lattice model
yields the first-order finite density phase transition at zero
temperature. These points are illustrated in 
Fig.~\ref{fig:muNtInfB} for $g_\sigma^2=2.0$.  For the tuned point 
${\bf T}(g_\sigma^2)$,  
for example, the left vertical phase boundary moves over the point 
between $\mu_L=0.12$ and 0.13. 
In Fig.~\ref{fig:f8} we translate the behavior into 
a schematic three-dimensional phase diagram 
for $N_T=\infty$ and a fixed value of $\mu_L$. 

\subsection{The Case of $\mu\neq 0$ and $T\neq 0$}   

\begin{figure}
\epsfxsize=0.9\textwidth
{\centering \epsfbox{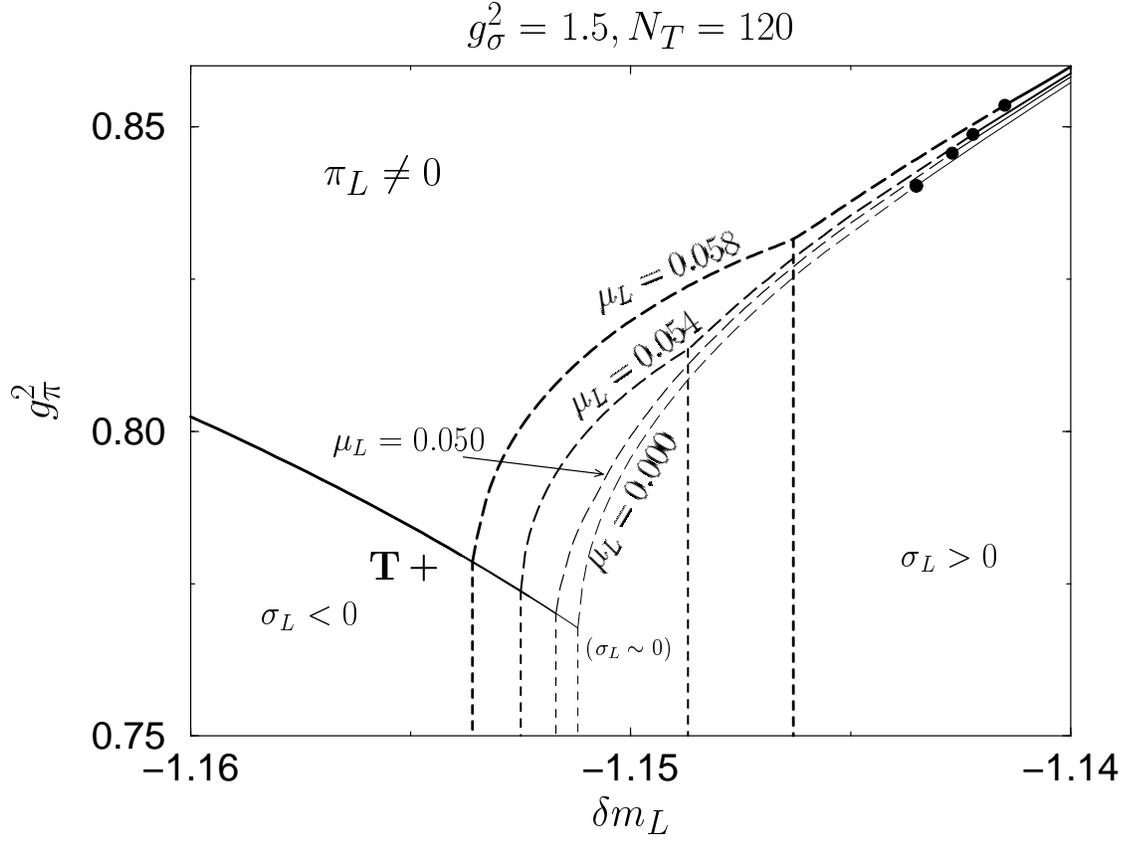} \par}

\vspace*{2cm}
\caption{Effect of \protect \( \mu _{L}\protect \) on the phase structure 
for large  \protect\( N_{T}\protect \).
\label{fig:NtMuPhase1st}}
\end{figure}

Finally we consider the finite-density transition at finite 
temperature. We examine this case by decreasing the temporal lattice size
from $N_T=\infty$.  The phase structure for $N_T=\infty$ found in
Sec.~\ref{subsec:fiveA} 
implies that the mechanism of generation of a first-order 
transition through a splitting of the $\sigma$ phase boundary as $\mu_L$ 
increases will remain operative for large temporal size $N_T$. 
We found this to be the case as shown in Fig.~\ref{fig:NtMuPhase1st} for 
$N_T=120$ at $g_\sigma^2=1.5$. 

\begin{figure}
\epsfxsize=0.9\textwidth
{\centering \epsfbox{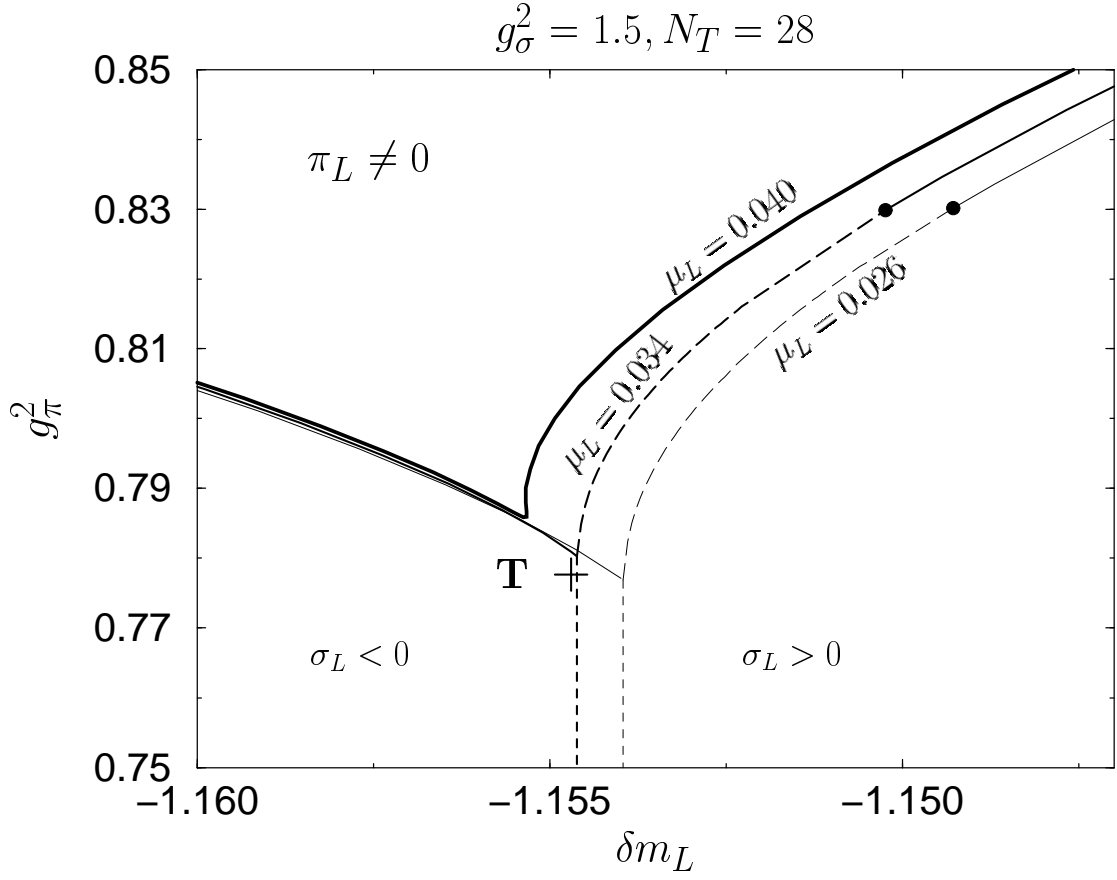} \par}

\vspace*{2cm}
\caption{Effect of \protect\( \mu _{L}\protect \) on the phase structure 
for small $N_T$. 
\label{fig:NtMuPhase2nd}}
\end{figure}

On the other hand, for small temporal sizes, 
we expect the behavior to become similar to the 
finite-temperature and zero chemical potential case examined in 
Sec.~\ref{sec:three}.  We plot an example of 
this case in Fig.~\ref{fig:NtMuPhase2nd} for which $N_T$ is decreased to 
28 : the parity-broken phase shrinks, and the first-order segment of the phase 
boundary and the $\sigma$ phase boundary disappear as one increases 
$\mu_L$.  In this case the phase transition is of second order 
in the continuum limit. 

\begin{figure}
\epsfxsize=0.9\textwidth
{\centering \epsfbox{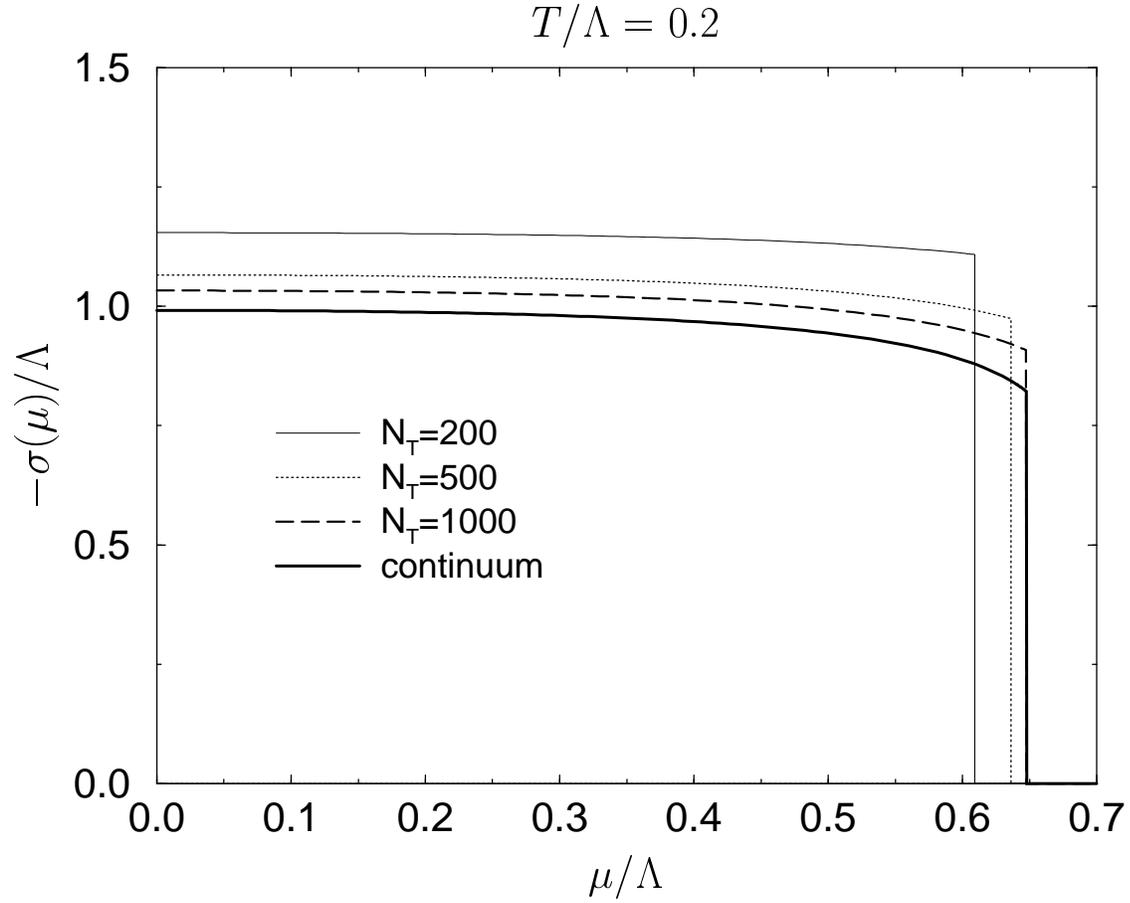} \par}

\vspace*{2cm}
\caption{\protect\( -\sigma \protect \) as a function of $\mu$ for 
$T/\Lambda=0.2$.  Couplings are tuned at ${\bf T}(g_\sigma^2)$.
\label{fig:sgNtMu1st1loop}}
\end{figure}

\begin{figure}
\epsfxsize=0.9\textwidth
{\centering \epsfbox{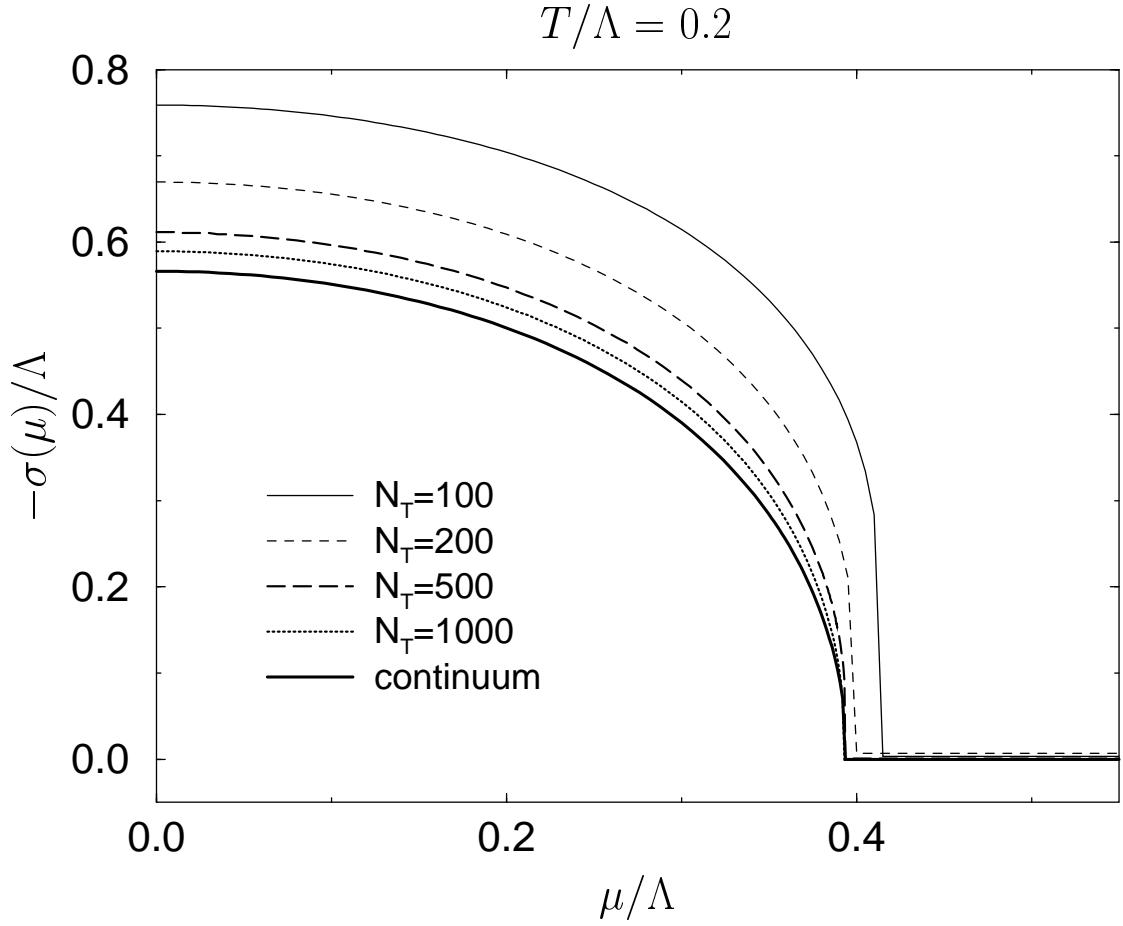} \par}

\vspace*{2cm}
\caption{Same as Fig.~\ref{fig:sgNtMu1st1loop} for $T/\Lambda=0.5$. 
\label{fig:sgNtMu2nd1loop}}
\end{figure}

We plot in Fig.~\ref{fig:sgNtMu1st1loop} and Fig.~\ref{fig:sgNtMu2nd1loop} 
curves of $\sigma(T, \mu)/\Lambda$ calculated as a 
function of $\mu/\Lambda$ for a fixed value of $T/\Lambda$ along the tuned
point ${\bf T}(g_\sigma^2)$. 
A clear discontinuity observed in Fig.~\ref{fig:sgNtMu1st1loop} for 
$T/\Lambda=0.2$ contrasts sharply with the convergence to second-order 
behavior seen for $T/\Lambda=0.5$ in Fig.~\ref{fig:sgNtMu2nd1loop}. 
This difference reflects the change of order of finite-density transition 
for large and small temporal lattice sizes discussed above. 
In the continuum limit we thus find a line of phase transition 
on the $(\mu, T)$ plane which 
switches from being of first-order to second order as the critical temperature 
increases. This agrees with the phase diagram of the continuum theory 
drawn in Fig.~\ref{fig:CLPhase}.

\section{Conclusions}
\label{sec:six}

In this article we have investigated the phase structure and the continuum 
extrapolation of the two-dimensional lattice Gross-Neveu model
at finite temperature and chemical potential. 
The choice of the Wilson action for fermions leads to 
the existence of a parity-broken phase separated from the parity-symmetric
phase by a two-dimensional boundary surface in the three-dimensional 
space of the couplings $(g_\sigma^2, g_\pi^2, \delta m_L)$ that specifies 
the model.  The phase transition across the surface 
is generally of second order.  However, in the region relevant for the 
continuum limit, ${\cal O}(a)$ effects of the Wilson term gives rise to 
a complicated structure, rendering the transition to be of first order in 
a region whose size vanishes in the continuum limit. 
The parity-symmetric phase is also divided into subregions by a first-order 
phase boundary corresponding to different minima of the effective 
potential.  We found that these detailed structures intertwine to lead to 
the continuum phase diagram reported in Ref.~\cite{Wolff}. 

An interesting issue is how much of the detailed structure we found for the
Gross-Neveu model has relevance for QCD in four dimensions.  
Of particular interest is the question that a part of the parity-flavor
breaking phase boundary may be of first order.  While the necessity of 
two independent couplings $g_\sigma^2$ and $g_\pi^2$ appears to be a 
specific feature of the Gross-Neveu model, arising from the presence of 
four-fermion couplings in the model, a distortion of the effective potential 
for effective meson fields in the $\sigma$ direction by terms odd in 
$\sigma$ would be present also for QCD, possibly triggering a first-order 
phase transition.  

The relevance of the structure we found for finite chemical potential is 
more likely; the emergence of three regions corresponding 
to three different values of $\sigma$ for large $\mu$ is a physical effect, 
not specifically tied with the use of the Wilson actions for fermions.  
The crossover from a first-order to second-order transition depending on 
the critical temperature means, however, that the order of the 
finite density transition may depend sensitively on the parameters of the 
model, perhaps the space-time dimension playing an 
important role\cite{Wolff}.  
These, then, are the issues which have to be directly addressed in QCD. 

\section*{Acknowledgments}

We thank Y. Taniguchi for informative discussions.
One of us (A.~ U. ) would like to thank F. Karsch for warm hospitality while 
visiting Zentrum f\"ur interdisziplina\"re Forschung (ZiF) of 
Bielefeld University under the program ``Multicritical 
phenomena in complex systems'', where part of the work was carried out.  
This work is supported in part by the 
Grants-in-Aid for Scientific Research from the Ministry of Education, 
Science and Culture (Nos. 2375, 10640246).  T.~I. is supported by Japan 
Society for the Promotion of Science.


\newpage
\newpage
\newpage
\newpage
\newpage
\newpage
\newpage
\newpage
\newpage
\newpage
\newpage
\clearpage
\newpage
\newpage
\newpage
\newpage
\newpage
\newpage
\newpage
\newpage
\end{document}